\documentclass[aps,prl,twocolumn,longbibliography,superscriptaddress]{revtex4-2}
\pdfoutput=1

\usepackage{amsmath}
\usepackage{amssymb}
\usepackage{graphicx}
\usepackage{float}
\usepackage{subfigure}
\usepackage{dcolumn}
\usepackage{bm}
\usepackage{bbm}
\usepackage{amsthm}
\usepackage{thmtools}

\usepackage{mathtools}
\usepackage{physics}
\usepackage{binarytree}
\usepackage{tikz,pgfplots}
\usepackage{verbatim}
\usepackage{pgfplots}
\usepackage{placeins}
\usepackage{algorithm2e}
\usetikzlibrary{decorations.pathreplacing}
\usetikzlibrary{arrows.meta}
\usetikzlibrary{graphs}
\usepackage{xcolor}
\usepackage{rotating}
\usepackage{yquant}
\usepackage{tabularray}
\usepackage{times}
\usepackage[colorlinks,bookmarks=true,citecolor=green,linkcolor=blue,urlcolor=blue]{hyperref}

\RestyleAlgo{ruled}
\pgfplotsset{compat=1.16}

\begin{document}

\title{Tangent Space Excitation Ansatz for Quantum Circuits}

\author{Ji-Yao Chen}
\email{chenjiy3@mail.sysu.edu.cn}
\affiliation{Center for Neutron Science and Technology, Guangdong Provincial Key Laboratory of Magnetoelectric Physics and Devices, School of Physics, Sun Yat-sen University, Guangzhou 510275, China}

\author{Bochen Huang}
\affiliation{Institute of Physics, Beijing National Laboratory for Condensed Matter Physics, Chinese Academy of Sciences, Beijing 100190, China}
\affiliation{School of Physical Sciences, University of Chinese Academy of Sciences, Beijing 100049, China}

\author{D. L. Zhou}
\affiliation{Institute of Physics, Beijing National Laboratory for Condensed Matter Physics, Chinese Academy of Sciences, Beijing 100190, China}
\affiliation{School of Physical Sciences, University of Chinese Academy of Sciences, Beijing 100049, China}

\author{Norbert Schuch}
\email{norbert.schuch@univie.ac.at}
\affiliation{University of Vienna, Faculty of Physics, Boltzmanngasse 5, 1090 Wien, Austria}
\affiliation{University of Vienna, Faculty of Mathematics, Oskar-Morgenstern-Platz 1, 1090 Wien, Austria}

\author{Chenfeng Cao}
\email{chenfeng.cao@connect.ust.hk}
\affiliation{HK Institute of Quantum Science $\&$ Technology, The University of Hong Kong, Hong Kong, China}
\affiliation{Dahlem Center for Complex Quantum Systems, Freie Universit\"{a}t Berlin, Berlin 14195, Germany}

\author{Muchun Yang}
\email{yang.muchun@iphy.ac.cn}
\affiliation{Institute of Physics, Beijing National Laboratory for Condensed Matter Physics, Chinese Academy of Sciences, Beijing 100190, China}
\affiliation{School of Physical Sciences, University of Chinese Academy of Sciences, Beijing 100049, China}

\date{\today}

\begin{abstract}

Computing excitation spectra of quantum many-body systems is a promising avenue to demonstrate the practical utility of current noisy quantum devices, especially as we move toward the ``megaquop'' regime. For this task, here we introduce a \textit{tangent space excitation ansatz} for quantum circuits, motivated by the quasi-particle picture of many-body systems and the structural similarity between quantum circuits and tensor networks. Increasing circuit depth by one layer to construct tangent space around the variational optimum of a parametrized quantum circuit, we show that massive low-energy single-particle states can be captured. Our ansatz relies on a distinct mechanism from that of excitation ansatz in matrix product state and projected entangled-pair state, and avoids intrinsic limitations of the latter.
Comparing our approach with existing quantum excited-state algorithms, we find that with similar computational cost, both the number of excited states and accuracy are significantly improved.
We demonstrate our ansatz in both one and two dimensions, and further show that this approach, implementable using Hadamard test, is scalable and suitable for current quantum processors.

\end{abstract}

\maketitle

\textit{Introduction.--}
Understanding low-energy properties of quantum many-body systems is of paramount importance in physics, with far-reaching applications in chemistry and materials science~\cite{Altland2010,Wen2004}. The underlying concept, quasi-particles, represents a hallmark of modern condensed matter physics. Classical approaches to this task face intrinsic challenges~\cite{Troyer2005,Orus2014}, while quantum computing~\cite{Nielsen2012}, a new computational paradigm, holds the promise of solving ground states of many-body systems without incurring the exponential wall or sign problem~\cite{Arute2019,Kim2023,Bluvstein2024,Acharya2025,Gao2025}. Since the early days of noisy intermediate-scale quantum era~\cite{Preskill2018}, the variational quantum eigensolver (VQE)~\cite{Peruzzo2014,Wecker2015,Yuan2019,Liu2019,McArdle2019,McArdle2020,Wiersema2020,Cerezo2021,Yalouz2021,Stanisic2022,Tilly2022,Bharti2022,Park2024,Cao2024a,Beseda2024,Tazi2024,Cianci2024} has been one of the most studied quantum algorithms that offers a potential pathway to quantum speedup. This promise has been extended to low-energy excited states, motivating the development of various quantum algorithms~\cite{McClean2017,Higgott2019,Nakanishi2019,Jones2019,Motta2020,Xu2023a,Tkachenko2024,Zhang2025,Farrell2025}. Nevertheless, locality, a key property of quantum many-body systems, has not been considered in its full strength in quantum algorithms for excited states, and massive excited states computation with high accuracy has not been achieved in quantum circuits.

Classically, with rapid progress in the last three decades, significant achievements have been realized in many-body physics using tensor network methods~\cite{Schollwock2011,Orus2019,Cirac2021}. Notably, with locality-based tangent space method, a generalization of the single mode approximation (SMA)~\cite{Feynman1954,Girvin1986,Arovas1988}, low energy excitation spectrum can be approached with matrix product state (MPS) for one dimension (1D)~\cite{Ostlund1995,Haegeman2012} and projected entangled-pair state (PEPS) for two dimensions (2D)~\cite{Vanderstraeten2019a,Chi2022,Ponsioen2023b,Tu2024}. However, the latter is limited to small bond dimension, due to intrinsic high computational cost. Given the structural similarity between tensor networks and quantum circuits, a new excitation ansatz based on locality may be possible for quantum computers, avoiding the classical computational cost. Yet, local entanglement degrees of freedom are not manifest in quantum circuits, making a simple adaption of excitation ansatz in MPS and PEPS unfeasible. Here, leveraging on the concept of tangent space, we propose an excitation ansatz for quantum circuits with distinct mechanism than that in MPS and PEPS, and demonstrate its broad applicability for quantum many-body systems. Targeting $O(N)$ excited states for $N$-qubit system, the measurement cost is at most quadratic in circuit depth and systems size, making it appealing to reach a few tens of qubits on current experimental platforms~\cite{Will2025}.


\textit{SMA and generalizations.--}
To motivate our ansatz, we start by discussing the single mode approximation method in many-body physics~\cite{Feynman1954,Girvin1986,Arovas1988}, taking 1D setting as an example.
Intuitively, as excited states in a free fermion system can often be described by adding particles to the ground state, for interacting system, one replaces particle with quasi-particle, which is a quantum of the collective variation over the ground state. For instance, the low-energy excitation of a ferromagnet is magnon, which is a mobile defect in a perfect alignment of magnetic moments and can also be viewed as quantum of the spin waves~\cite{Auerbach1994}. Concretely, given a ground state $|\Psi\rangle$ of a $N$-site translation invariant system, the SMA constructs an excited state of momentum $K$ by perturbing $|\Psi\rangle$ with a momentum superposition of onsite operator $G$: $|\Phi_K(G)\rangle = \sum_{j=0}^{N-1}\mathrm{e}^{-\mathrm{i}jK}G_j|\Psi\rangle$, which with optimized $G$ has reasonable accuracy for low-energy single particle excitations in gapped systems. In the case of magnon, $G$ could be a spin flip operator. Here normalization of $|\Phi_K(G)\rangle$ is left to the associated variational optimization and $K$ is in fact momentum of the quasi-particle, taking a value $K=2\pi m/N (m=0,1,\ldots,N-1)$. 
Motivated by Ref.~\cite{Haegeman2013a}, one can further enlarge the spatial support of local perturbations (denoted as $n$ hereafter) to lower the approximation error, especially for isolated bands in many-body spectrum.

Applying the idea of SMA to MPS yields a specific form of excitation ansatz.
As operators with large support are not convenient to handle, one instead allows perturbations to act on both physical and entanglement degrees of freedom of a local tensor. The effect of the latter can spread to a region with size on the scale of correlation length, effectively realizing operators with large spatial support on the physical degrees of freedom. Since only a single local tensor is perturbed, one can view the excited state as living in the tangent space of the MPS manifold~\cite{Vanderstraeten2019b}. Similarly for PEPS.

Likewise, SMA also suggests an excitation ansatz for quantum circuits. As $|\Phi_K(G)\rangle$ depends on $G$ linearly, one can take $G$ from operator basis of single qubit $\{X, Y, Z, I\}$, with Pauli matrices $X,Y,Z$ and identity operator $I$. The resulting four states form a subspace for the low-energy states.
This is indeed known as subspace expansion VQE (SEVQE) for excited state~\cite{McClean2017}.
However, the generality of SMA and range of excited states captured are typically limited (see examples later), and to improve it, one needs to consider operators with larger support. Unlike MPS, where perturbations on entanglement degrees of freedom can spread, one does not have direct access to compact local tensors of the state on quantum devices. Therefore, a fundamentally new way of effectively realizing operator with large spatial support is needed.


\textit{Excitation ansatz for quantum circuits.--}
On quantum computers, the ground state $|\Psi\rangle$ of an $N$-qubit system with Hamiltonian $H$ is generated by a depth $D$ unitary circuit acting on an easy-to-prepare state $|\Psi_0\rangle$: $|\Psi\rangle=U_D U_{D-1}\cdots U_1 |\Psi_0\rangle$. Here, $U_l$ on the $l$th layer is composed by a set of short-range gates acting on non-overlapping qubits, which can be variationally determined using various hybrid quantum-classical algorithms~\cite{Tilly2022,Schiffer2022}. We further assume that $|\Psi\rangle$ is translationally invariant, which can be realized with properly chosen unitary gates. For the case without explicit translation invariance, see End Matter and supplemental material (SM)~\footnote{See the Supplemental Material [url], in which we compare our approach with several existing quantum excited-state algorithms, show fermion solution~\cite{Mbeng2024} and additional data for 1D TFI model, and describe the HVA and supplementary data for 2D TFI model. We further discuss the HVA for kagome Heisenberg model and details of excitation and dynamical spin structure factor for a Heisenberg chain.}.

\begin{figure}[htbp]
\centering
    \includegraphics[width=0.95\columnwidth]{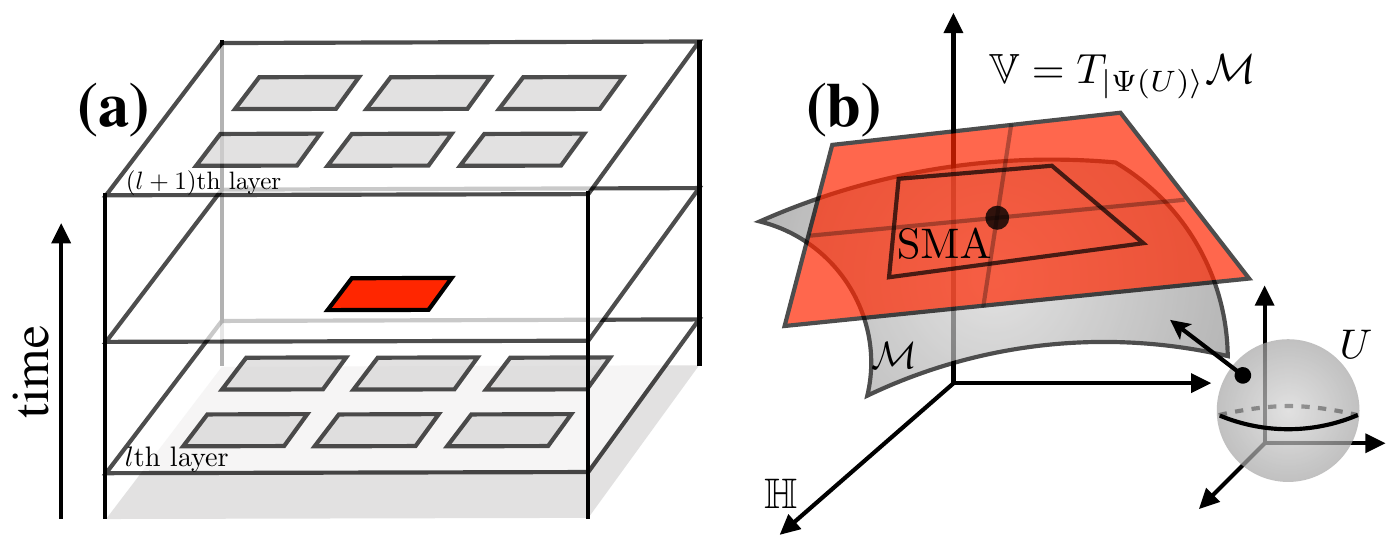}
\caption{Schematics for the excitation ansatz. (a) Adding a perturbation (shown in red) at arbitrary layer of the ground state circuit generates one configuration of the low-energy subspace. (b) The excitation ansatz is in the tangent space $T_{|\Psi(U)\rangle}\mathcal{M}$ of a manifold $\mathcal{M}$ in the full Hilbert space $\mathbb{H}$, which incorporates the SMA as a subspace.}
\label{fig:excitation_ansatz}
\end{figure}

As the basis of excitation in SMA is obtained by adding a perturbation layer on the final variational ground state, to go beyond SMA, one reasonable way is to insert the perturbation at arbitrary layer of the ground state circuit, see Fig.~\ref{fig:excitation_ansatz}(a) for a schematic illustration. Moreover, the single site operator $G$ in SMA can be naturally extended to a multisite gate. With this general scheme, we now describe our ansatz.

Suppose the added gate $G$ (located between layer $l$ and $l+1$) acts non-trivially on $n$ qubits, which is taken from the operator basis $\{X,Y,Z,I\}^{\otimes n}$. For a momentum $K$ sector, we take a superposition to form the state
\begin{equation}
    |\phi(G)\rangle= \sum_{j=0}^{N-1} e^{-\mathrm{i}j\frac{2\pi m}{N}}T^{j} U_D\cdots U_{l+1} G U_{l}\cdots U_1|\Psi_0\rangle,
\label{eq:1D_basis}
\end{equation}
which is an unnormalized eigenstate of translation operator $T$ with momentum $K=2\pi m/N$ ($m=0,1,\ldots,N-1$). (Here for notational simplicity, we introduce the state in 1D setting.) Then we collect states with all different $G$ to construct a subspace $\mathbb{V}_l$ of the total Hilbert space $\mathbb{H}$. We can further obtain a space $\mathbb{V}$ which is a direct sum of $\mathbb{V}_l$ with different $l$: $\mathbb{V}=\oplus_l\mathbb{V}_l$. It is clear that the SMA is included in $\mathbb{V}$ (see Fig.~\ref{fig:excitation_ansatz}(b)), and is systematically improved by subspaces $\mathbb{V}_l$ with smaller $l$, providing a physical intuition that $\mathbb{V}$ is close to the low-energy sector. As the dimension of each $\mathbb{V}_l$ is independent of system size $N$, $\mathbb{V}$ has a dimension linear in circuit depth $D$, while independent of $N$.

Our ansatz is defined in $\mathbb{V}$. Denoting the total number of states Eq.~\eqref{eq:1D_basis} as $\chi$, states $\{|\phi_j\rangle, j=1,2,\ldots,\chi\}$ form a set of overcomplete basis for $\mathbb{V}$. The variational excited state takes a form $|\Phi\rangle=\sum_{i=1}^{\chi} c_i|\phi_i\rangle$ with parameters $\mathbf{v}=(c_1,c_2,\ldots,c_{\chi})^{\mathrm{T}}$. Since $|\Phi\rangle$ depends linearly on $\mathbf{v}$, the optimization of $\mathbf{v}$ can be achieved via solving a generalized eigenvalue equation: $\mathbf{H}\mathbf{v}=E\mathbf{N}\mathbf{v}$. Here $\mathbf{N}$ is the norm matrix in $\mathbb{V}$ while $\mathbf{H}$ is the effective Hamiltonian: $\mathbf{N}_{ij}=\langle\phi_i|\phi_j\rangle$, $\mathbf{H}_{ij}=\langle\phi_i|H|\phi_j\rangle$. Given possible zero modes in $\mathbf{N}$, one takes a pseudo-inverse of $\mathbf{N}$ (denoted as $\mathbf{N}^{-1}$) and diagonalizes $\mathbf{N}^{-1}\mathbf{H}$, through which the variational energy $E$ and eigenvector $\mathbf{v}$ are found.

Fixing $n$, here we discuss the computational cost of our algorithm for 1D systems on quantum devices. Similar analysis holds for 2D case. An essential factor is circuit depth $D$ of variational ground state, which depends on the circuit architecture. In this work we have used the Hamiltonian variational ansatz (HVA) for ground states~\cite{Wecker2015,Ho2019,Wiersema2020}, which has a linear depth $D\sim O(N)$. When implementing the ansatz, besides using linear combination of unitary technique~\cite{Childs2012} to implement Eq.~\eqref{eq:1D_basis} in one circuit, one can consider each configuration with fixed position of $G$ separately, so that full dispersion relation of the quasi-particle can be obtained simultaneously through classical post processing. 
We analyze the latter case in the following. With HVA, the dimension of space {\it{with all momentum sectors}} scale as $O(DN)\sim~O(N^2)$. To obtain the norm matrix $\mathbf{N}$ and effective Hamiltonian $\mathbf{H}$, one can use the Hadamard test to measure each individual matrix element (see End Matter), a specialized instance of which was recently realized in quantum devices~\cite{Will2025}. Using translation symmetry, the measurement cost for $\mathbf{N}$ is $O(D^2N)\sim O(N^3)$, while for $\mathbf{H}$ it is $O(D^2N^2)\sim O(N^4)$. After Fourier transform, the two matrices for each momentum sector can be obtained, each of which has a dimension $O(D)\sim O(N)$, and can be diagonalized efficiently on classical computers with a cost $O(D^3)\sim O(N^3)$. Note that, for part of the matrix elements, using cancellations with unitary, the Hadamard test can be replaced with standard measurement of correlations, reducing run-time. Overall, with above conservative estimate, application of our ansatz with a few tens of qubits would be within reach of current devices~\cite{Will2025}.


\textit{Tangent space of quantum circuit.--} 
Above is the framework of our excitation ansatz. As already mentioned, generalization of SMA for MPS and PEPS leads to the concept of tangent space. We now analyze the structure of our ansatz from this geometric point of view.

Taking each gate of the quantum circuit as an independent local tensor, 
a variational manifold $\mathcal{M}$ is formed by the resulting tensor network state. At the variational optimum, a tangent space $T_{|\Psi(U)\rangle}\mathcal{M}$ can be constructed by taking derivative with respect to each tensor (see Fig.~\ref{fig:excitation_ansatz}(b)), with basis generated in the following way. Consider $i$th gate in $l$th layer of the circuit $U_{l,i}$, we replace this gate by a new gate $G$ from the corresponding operator basis while keeping all other gates fixed.
We can further enumerate all gates in the ground state circuit, do above gate replacement, and collect resulting states together, forming a set of (overcomplete) basis for the tangent space $T_{|\Psi(U)\rangle}\mathcal{M}$ of the variational manifold $\mathcal{M}$. Now, using $G=G'U_{l,i}$ one can see that adding one layer with local gate is equivalent to above gate replacement
~\footnote{One caveat is that the support of $G$ and $G'$ could be slightly different.}. Then taking a direct sum of all momentum sectors, it is easy to see that our excitation ansatz Eq.~\eqref{eq:1D_basis} is in fact a Fourier transform in the tangent space. Thus we have 
\begin{equation}
\mathbb{V}=T_{|\Psi(U)\rangle}\mathcal{M}\equiv\{G\partial_U |\Psi(U)\rangle\}.
\label{eq:tangent_space}
\end{equation} 
One advantage of adding one layer is that translation symmetry can be kept explicit, and size of support of added gate can be taken as a hyperparameter of the ansatz.

Although sharing the tangent space structure, the underlying mechanism of our ansatz is distinct from that in MPS. 
Motivated by Ref.~\cite{Haegeman2013a}, we analyze how perturbations with large spatial support, crucial for reducing approximation error, are approached in both cases.
For excitation ansatz in MPS, virtual degrees of freedom can spread local perturbation. On the contrary, only physical degrees of freedom are available in quantum circuits.
However, apart from the fact that $\mathbb{V}$ incorporates and improves upon SMA, we notice that, for an added gate away from the top layer, the effect of this gate can propagate through subsequent layers of the circuit, forming a light cone~\cite{vonKeyserlingk2018,Nahum2018}. E.g., taking gate
added between layer $l+1$ and $l$, which generates state $|\phi\rangle = U_D\cdots U_{l+1}GU_{l}\cdots U_1|\Psi_0\rangle$, $|\phi\rangle$
can be equivalently written as $|\phi\rangle = \tilde{G}|\Psi\rangle$ with $\tilde{G} = U_D\cdots U_{l+1} G U_{l+1}^\dagger \cdots U_D^\dagger$. This suggests that local perturbation in deep layers can
spread over a large region on the lattice, effectively realizing a structured subset of operators with large spatial support.

Besides confronting our approach with MPS and Ref.~\cite{Haegeman2013a}, for heuristics guided circuits, we can also understand the effect of operator spreading from a more physical way. 
Taking the HVA~\cite{Wecker2015,Ho2019,Wiersema2020} as an example, which mimics adiabatic evolution, an excitation in a deep layer is close to the renormalization group fixed point, where a single particle excitation is indeed localized. This excited state is then adiabatically evolved by the HVA circuit, transforming it to the excitation for the final Hamiltonian, in particular for an isolated band of excitations. Even for a state in the continuum, a reasoning along these lines is still plausible, given the excitations are variationally optimized.

Before moving further, we would like to mention that this ansatz is based on the single quasi-particle picture, and thus not expected to capture the full continuum spectrum with many-particle excitations. The latter can be approximated by taking a linear combination of single-particle dispersion relation. For a more precise description, a two-particle ansatz has been proposed in MPS~\cite{Vanderstraeten2015b} and is beyond the scope of this work.
In the following we will numerically demonstrate in both 1D and 2D that this approach represents a state-of-the-art advancement 
for quantum excited-state algorithm.


\begin{figure}[h]
\centering
\includegraphics[width=0.98\columnwidth]{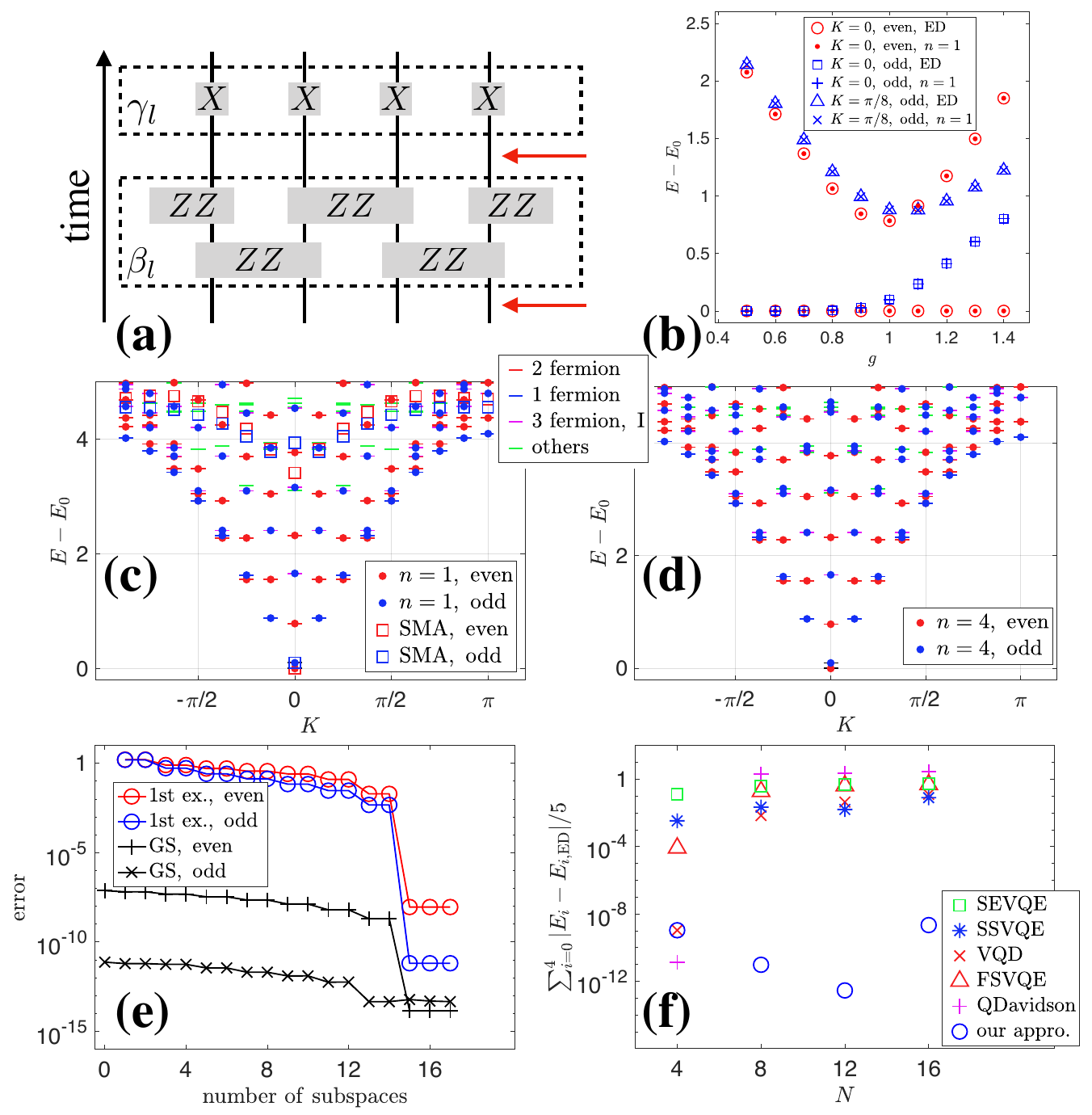}
\caption{Tangent space excitation for 1D TFI model. (a) HVA for ground state. Red arrows indicate possible locations of perturbation layer. (b) A comparison between lowest four energy levels of a $16$-site system obtained via the $n=1$ excitation ansatz and ED, showing high accuracy of the ansatz. (c) Zooming in $g=1$, the $n=1$ ansatz shows a better performance than SMA and
captures one-particle excitations, i.e. ``2 fermion'', ``1 fermion'', and ``3 fermion, I'' states in the exact solution (shown in SM), while missing many-particle excitations. The latter can be captured with $n=4$, shown in (d). In (e), at $g=1$, 
with gradually adding subspaces in deeper layers to SMA, the accuracy is systematically improved.
With all layers included, the error of excited states are on the same order of HVA ground state. We further compare our $n=1$ ansatz with existing methods for excited states in (f), finding several orders of magnitude improvement in accuracy. The varying accuracy with $N$ in our approach is due to fluctuations in HVA ground state optimization.}
\label{fig:1D_Ising}
\end{figure}

\textit{Numerical results in 1D.--}
The first model we consider is the 1D transverse field Ising (TFI) model with Hamiltonian: $H_{\text{1D}} = -\sum_{i=1}^{N} Z_iZ_{i+1}+gX_i$, where $g>0$ and periodic boundary condition (PBC) is imposed. This model has a $\mathbb{Z}_2$ parity symmetry with symmetry operator $U_{\mathbb{Z}_2}=\prod_i X_i$, which spontaneously breaks when $g<1$. The spectrum can be labeled by parity and momentum quantum numbers.

We adopt the HVA~\cite{Ho2019,Wiersema2020} shown in Fig.~\ref{fig:1D_Ising}(a) for the ground state, with $ZZ$ gate $U_{ZZ}=\mathrm{exp}(-\mathrm{i}\beta_l Z_iZ_{i+1}/2)$ for nearest neighbor qubits $i,i+1$ and $X$ gate $U_X=\mathrm{exp}(-\mathrm{i}\gamma_l X_i/2)$ for qubit $i$, which are parameterized by $\beta_l,\gamma_l$, respectively. Here $\beta_l,\gamma_l\in\mathbb{R}$, $l=1,2,\ldots,N/2$. Note that, the $ZZ$ gates in two consecutive layers share the same parameter to keep translation symmetry explicit.
For this model, it is useful to first target the lowest energy state of each parity sector separately. 
For parity even sector, we use a initial state $|\Psi_0\rangle = |+\rangle^{\otimes N}$ (here $X|+\rangle=|+\rangle$), while for the odd sector, a Greenberger–Horne–Zeilinger (GHZ) type state $|\Psi'_{0}\rangle=(|0\rangle^{\otimes N} - |1\rangle^{\otimes N})/\sqrt{2}$ is used. Then we optimize the HVA circuit in each case independently, where a circuit depth $D=3N/2$ ($N$ even) is known to be sufficient~\cite{Ho2019}.

Applying the excitation ansatz (Eq.~\eqref{eq:1D_basis}), with added layer located between $ZZ$ layer and $X$ layer (see Fig.~\ref{fig:1D_Ising}(a)), we obtain
the low-lying spectrum of this model. {With minimally inserting a $n=1$-site gate to the ground state circuit, through comparing with exact diagonalization (ED), we find the accuracy of even (odd) parity spectrum from even (odd) parity ground state is considerably better than the opposite parity spectrum (see SM for data).}
Taking even (odd) parity spectrum with even (odd) ground state,
we arrive at the low energy level spectroscopy of 1D TFI model, showing a perfect agreement with ED for a wide range of $g$. See Fig.~\ref{fig:1D_Ising}(b) for the $N=16$ case.

Focusing on $g=1$, Fig.~\ref{fig:1D_Ising}(c) shows that our $n=1$ ansatz can capture significantly more states with higher accuracy than SMA, where both approaches use same ground states.
Through comparison with exact fermion solution, Fig.~\ref{fig:1D_Ising}(c) further shows that
the $n=1$ ansatz can describe best the one-particle excitation while missing many-particle excitations. The latter can be captured by increasing $n$, as shown in Fig.~\ref{fig:1D_Ising}(d) with a $n=4$ ansatz. This high accuracy for one-particle states remains stable with increasing system size, suggesting our ansatz is scalable for this case (see SM for data).
However, we also find that for $N=8,12$, the $n=3$ ansatz is sufficient to capture multi-particle states (data not shown), reversely suggesting that increasing $n$ for larger sizes may not be tractable. A detailed investigation is left to future works.

To quantify the effect of perturbations in deep layers, we define a sequence of space by gradually adding $\mathbb{V}_l$ with smaller $l$ into the variational space and the first space is the SMA.
Taking the mean error of the first excitations in all momentum sectors as a measure, we compare the accuracy of this sequence with $n=1$. As shown in Fig.~\ref{fig:1D_Ising}(e),
with $g=1$, the error of SMA is indeed large, and including deeper layers can systematically improve the accuracy. With space $\mathbb{V}$, the error is essentially on the same order of HVA ground state, demonstrating that deep layers are crucial for the variational power.
We note in passing that the accuracy of ground state is also improved with our ansatz.

To further evaluate the performance, we compare our approach with several existing quantum excited-state algorithms, including SEVQE~\cite{McClean2017}, subspace search variational quantum eigensolver (SSVQE)~\cite{Nakanishi2019}, variational quantum deflation (VQD)~\cite{Higgott2019}, folded spectrum VQE (FSVQE)~\cite{Tazi2024}, and quantum Davidson (QDavidson) algorithm~\cite{Tkachenko2024}.
Here the figure of merit is the error of lowest five energy eigenvalues of 1D TFI model at $g=1$. As shown in Fig.~\ref{fig:1D_Ising}(f), with moderate system size $N$, our approach significantly outperforms existing methods, reducing the error by several orders of magnitude.
Details of the comparison and a discussion of related algorithms~\cite{Yalouz2021,Beseda2024,Cianci2024} are provided in the SM.


\begin{figure}[h]
\centering
\includegraphics[width=0.95\columnwidth]{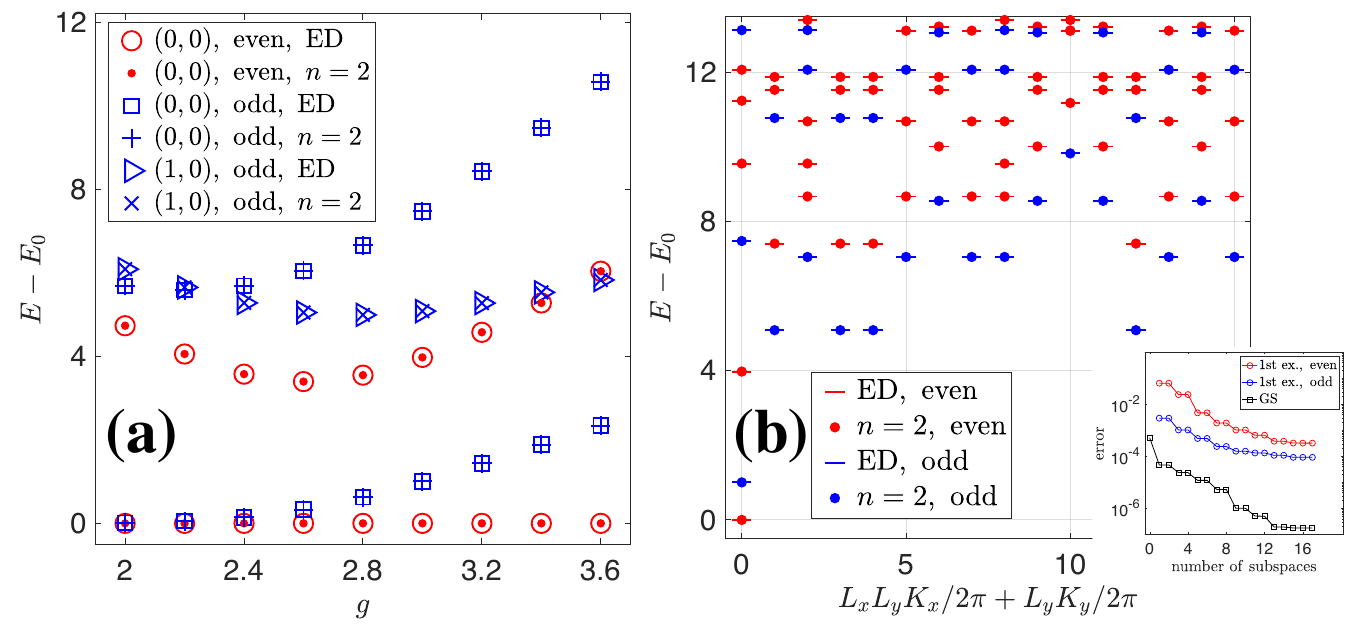}
\caption{Tangent space excitation for 2D TFI model on a $N=4\times 4$ square lattice.
(a) We compare the lowest five levels with ED, finding good agreement across the phase transition. (b) We show with an example at $g=3$ (close to the critical point) that 
massive low-energy levels can be captured. In the inset of (b), we show that average errors of parity resolved first excitations in all momentum sectors decrease with increasing number of subspaces $\mathbb{V}_l$, reaching values in the same order as HVA ground state error.
}
\label{fig:2D_Ising}
\end{figure}

\textit{Numerical results in 2D.--}
Our ansatz can be easily extended to 2D and higher dimension, showing promise in probing excitation spectrum of challenging systems. Indeed, the PEPS based excitation ansatz 
is severely hindered by the high computational cost and involved tensor diagrams, limiting it to small bond dimension. See, e.g., Ref~\cite{Chi2022}. In contrast, entanglement is not a restriction on quantum circuits, and thus circuit based excitation ansatz is likely to provide more accurate results for high dimensional systems.

Consider a square lattice with linear size $L_x(L_y)$ and translation operator $T_x(T_y)$ in the $x(y)$ direction. Assuming translation symmetry is kept explicitly,
the basis of the ansatz is constructed in the same way as Eq.~\eqref{eq:1D_basis}, taking a form $|\phi_{\bf K}(G)\rangle=\sum_{j_2=0}^{L_y-1}\sum_{j_1=0}^{L_x-1}e^{-\mathrm{i}(j_1K_x + j_2K_y)}T_y^{j_2}T_x^{j_1}\cdot U_{D}\ldots U_{l+1}G U_{l}\ldots U_1|\Psi_0\rangle$, which is an unnormalized eigenstate of translation operators with eigenvalues ${\bf K}=(2\pi m_1/L_x, 2\pi m_2/L_y)$, where $m_1=0,\ldots,L_x-1, m_2=0,\ldots,L_y-1$. 
When considering a
multi-qubit gate $G$, $G$
can have orientations. E.g., a nearest-neighbor gate can act on $x$ or $y$ bond. It is then reasonable to include both types of basis states in the space.

With this setup, we show results for 2D transverse field Ising model on the square lattice. The Hamiltonian reads $H_{2D}=-\sum_{\langle i,j \rangle}Z_iZ_j - g\sum_{i}X_i$, where the first sum runs over all nearest-neighbor bonds and PBC is considered. In the thermodynamic limit, this model is known to have a critical point with $g_c=3.04$~\cite{Blote2002}, separating a $\mathbb{Z}_2$ symmetry breaking phase at small $g$ and a symmetric phase at large $g$.

Following 1D, we use a similar HVA ansatz for the 2D TFI model~\cite{Cao2025Unveiling}, where unitary gates in each layer follow the geometric locality of the lattice. The system size we consider is $N=4\times 4$, and for all parameters $g$, we use a depth $D=5N/2$ circuit. With no evidence for exact ground states on quantum circuits, we variationally optimize circuit parameters to reach a ground state energy error around $1\times (10^{-3}-10^{-4})$, and then use the 
excitation ansatz to compute low-lying spectrum. We note that, unlike 1D, GHZ type initial state turns out to have no superior performance than initial state $|+\rangle^{\otimes N}$, except for small $g$ where the ground state is close to a GHZ state. Here
we will only consider initial state $|+\rangle^{\otimes N}$ with the $n=2$ ansatz, which
has higher accuracy than the $n=1$ case (see SM for details).

In Fig.~\ref{fig:2D_Ising}(a), we compare the tangent space excitation spectrum with ED for a wide range of $g$.
On both sides of the phase transition, the low-energy spectrum can be accurately captured. Taking a closer look at the spectrum, we find that, in addition to the lowest excitation in each momentum sector, many of the low-energy excited states can be approximated, as shown in Fig.~\ref{fig:2D_Ising}(b). Note that, below $g_c$,
the low-energy spectrum contains both one-magnon excitation and two-magnon bound state~\cite{Dusuel2010,Ponsioen2020}, suggesting our ansatz can capture this type of excitations.
In the inset of Fig.~\ref{fig:2D_Ising}(b), we show how the error decreases with increasing number of subspaces $\mathbb{V}_l$, starting from the top layer, where SMA is a subspace. 
Similar to 1D, excited states reach the same level of accuracy as HVA ground state, though the behavior is more smooth. Detailed comparisons with SMA are in SM.


\textit{Discussion and outlook.--}
We have shown that our approach, with mechanism distinct from that of MPS and PEPS, significantly outperforms known methods, and is scalable and suitable for current devices.
In the End Matter we show applications for spectrum of 2D kagome Heisenberg antiferromagnet and dynamical structure factor of a spin-$1/2$ Heisenberg chain, bringing useful quantum advantage within reach.

An interesting further question is about the aforementioned multi-particle excitation.
For that, a potentially scalable way is to consider multiple perturbations separated by certain distance and time, with which a space can still be constructed. On the other hand, one may wonder whether the barren plateau (BP) phenomena~\cite{McClean2018Barren} would hinder applications of this ansatz. While we have mostly used the HVA in this work, the scheme can be generalized to other circuit layouts, where BP phenomena can be mitigated~\cite{Zhang2024, Cao2024Exploiting}. It would be interesting to study how the performance of the excitation ansatz depends on the circuit structure,
e.g., brickwall, sequential~\cite{Wei2022}, parallel sequential~\cite{Wei2025}, finite local depth circuit~\cite{Zhang2024}, and circuits with measurement, feedforward and control~\cite{Piroli2021, Iqbal2024, Tantivasadakarn2024, Chen2025Nishimori, Cao2025Measurement, Zi2025, Yan2025}, 
and see how the accuracy varies. We leave these questions to future study.


\begin{acknowledgments}
\textit{Acknowledgments.--}
We thank Frederik Wilde, Zheng An, Guang-Ming Zhang, Tao Xiang, Sylvain Capponi, Laurens Vanderstraeten for discussions, and Matthieu Mambrini for help with figures. This work is supported by National Natural Science Foundation of China (Grants No.~12447107, No.~12304186), National Key Research and Development Program of China (Grants No.~2021YFA0718302 and No.~2021YFA1402104), Guangdong Basic and Applied Basic Research Foundation (Grant No.~2024A1515013065), Guangzhou Basic and Applied Basic Research Foundation (Grant No.~2024A04J4264), the Austrian Science Fund (FWF) via Grants~\href{https://doi.org/10.55776/COE1}{10.55776/COE1} and~\href{https://doi.org/10.55776/F71}{10.55776/F71}, the European Union -- NextGenerationEU, the European Union’s Horizon 2020 research and innovation programme through Grant No.\ 863476, and the Alexander von Humboldt Foundation.
\end{acknowledgments}

\textit{Data and code availability.--} Data, data analysis, and simulation codes are available upon reasonable request on Zenodo~\cite{zenodo}.

\bibliography{draft}

\clearpage
\newpage
\appendix
\renewcommand\thefigure{E\arabic{figure}}  
\renewcommand{\theequation}{E\arabic{equation}}
\renewcommand\appendixname{EM}
\setcounter{figure}{0} 
\setcounter{equation}{0}


\section*{End Matter}

\section{Further applications}
\label{sec:application}

\begin{figure}[h]
\centering
\includegraphics[width=0.98\columnwidth]{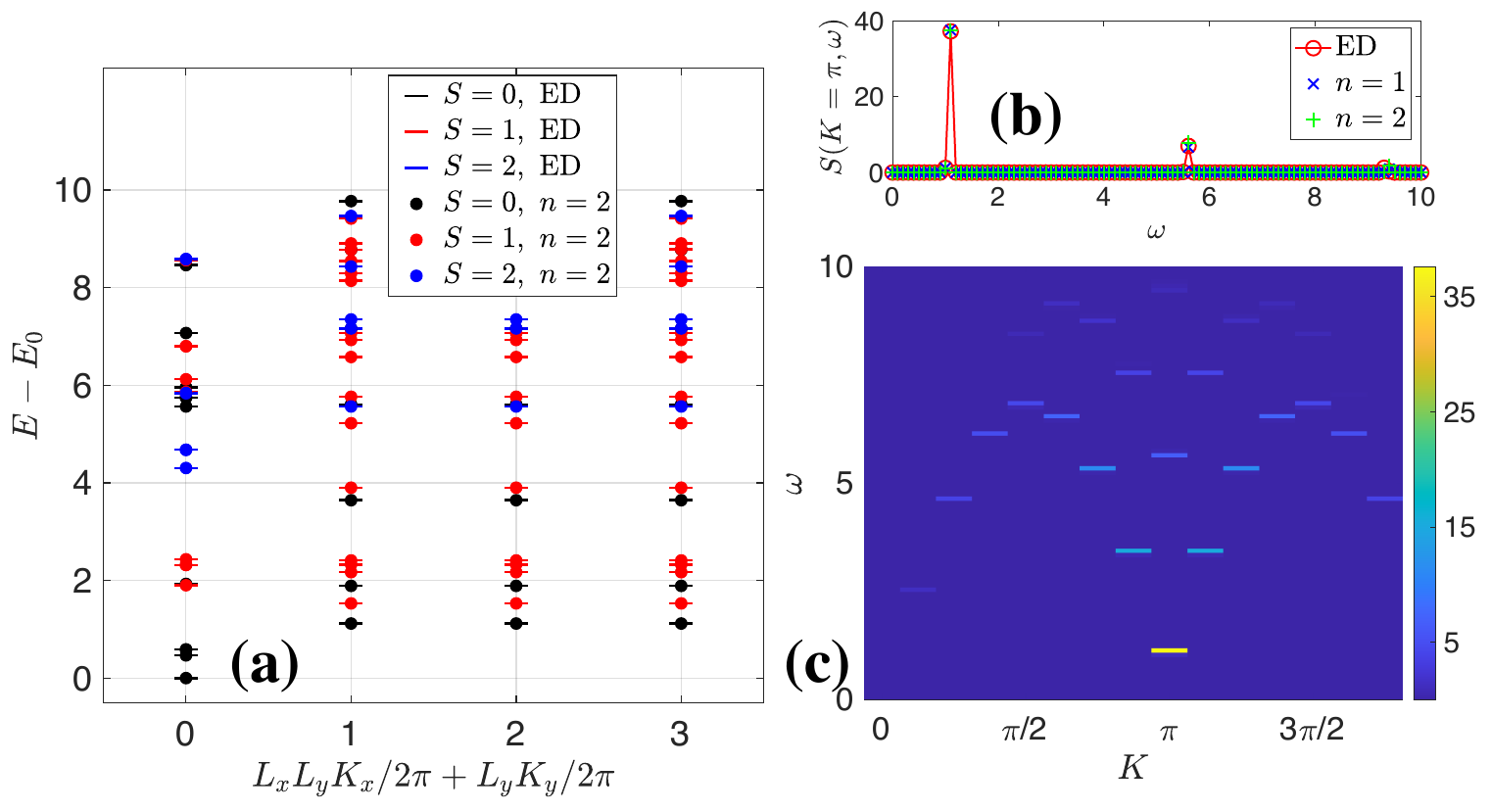}
\caption{Application of the excitation ansatz. In (a), we compare the excitation spectrum obtained using a $n=2$ ansatz with ED, for a $L_x=L_y=2$ kagome lattice. 
The degeneracy due to $\mathrm{SU}(2)$ symmetry and lattice symmetry is perfectly recovered. Here $K_x$ and $K_y$ correspond to momentum along two translation directions. In (b),(c), we compute the DSF of a $16$-site spin-$1/2$ Heisenberg chain, using the excitation ansatz. The $n=1$ ansatz captures triplet excitation, which is the excitation captured by DSF due to selection in quantum number. In (b), we compare the DSF at $K=\pi$ with ED, finding good agreement. (c) shows a full scan of DSF computed with $n=1$ ansatz, which agrees with ED (shown in the SM).}
\label{fig:application}
\end{figure}

To further demonstrate the usefulness of our ansatz and potential quantum advantage, we now consider applications of the excitation ansatz to systems and physical observables that are challenging for classical methods. Note that, in these systems, due to symmetry reason, it is not easy to keep translation symmetry explicit. We will 
show that in these cases, the tangent space excitation can be easily adapted and is as useful as in the translationally invariant case.

The first application is the spin-$1/2$ kagome Heisenberg antiferromagnet (KHA), which is known to host a quantum spin liquid ground state with the precise nature under intense debate~\cite{He2017,Liao2017,Mei2017}. One potential way of resolving the spin liquid nature is to consider the excitation spectrum~\cite{He2017}. The Hamiltonian of this system is given by $H_{\mathrm{KHA}}=\sum_{\langle i,j \rangle} X_iX_j + Y_iY_j + Z_iZ_j$, where the sum runs over all nearest-neighbor pairs on the kagome lattice, with 
PBC in both directions imposed. See the SM for illustration of the cluster and boundary condition used in this work.

The KAH model has a global $\mathrm{SU}(2)$ spin rotation symmetry, and the ground state on finite torus is known to be in the spin singlet sector~\cite{Lauchli2019}. Early studies~\cite{Kattemolle2022,Bosse2022} suggest a HVA for the ground state of this model, with a Heisenberg gate $U(\beta)$ acting on each of the nearest-neighbor bonds on the lattice. Here 
$U(\beta) = \mathrm{exp}\big(-\text{i}\beta/2 (X_iX_j+Y_iY_j+Z_iZ_j)\big )$ for a 
bond connecting site $i$ and $j$. Notice that, the ground state of a single nearest-neighbor term is the singlet state $|\Psi^-\rangle = (|01\rangle - |10\rangle)/\sqrt{2}$. To ensure the variational state is a global spin singlet, in our simulations, we use a valence bond covering of the lattice as the initial state, where each valence bond represents a spin singlet. In addition, we take each $\beta$ 
being independent, 
and refer to the SM for further details of the variational ground state simulation.

Since the initial state and the circuit for the ground state is not translationally invariant, we do not impose translation symmetry in the basis for excitation. Instead, we collect all states where a single gate is inserted in arbitrary 
location of the ground state circuit to form the space, where we then diagonalize the Hamiltonian. The momentum quantum number of excited state is computed afterwards and we further focus on the total spin quantum number. The result for a $12$-site kagome cluster with circuit depth $D=64$ is shown in Fig.~\ref{fig:application}(a), showing a good agreement with the ED results. Note that, since a $n=1$ ansatz can only capture the 
spin triplet excitation, one needs to consider at least a $n=2$ ansatz to reveal the excitations with other spin quantum numbers.

With excited states at hand, another application is to study the dynamical spin structure factor (DSF) of a quantum many-body system, which can be measured by neutron scattering experiments. Here we consider the spin-$1/2$ Heisenberg chain with Hamiltonian $H_{\mathrm{chain}}=\sum_{i} X_iX_{i+1}+Y_iY_{i+1}+Z_iZ_{i+1}$, with PBC imposed. We use a HVA to get the singlet variational ground state, and further use excitation ansatz to compute the excited state. Here due to the $\mathrm{SU}(2)$ symmetry we only consider the $z$ component of the DSF, which is defined as:
$
S(K,\omega)=\sum_a |M_K|^2\delta(\omega-E_a+E_0),
$
with $M_K=|\langle \Phi_a|Z_K|\Psi\rangle|$ and $E_0,E_a$ being the energy of ground state and $a$th excited state, respectively. Notice that $Z_K$ is the Fourier transform of the spin operator: $Z_K=\frac{1}{\sqrt{N}}\sum_j \mathrm{e}^{-\mathrm{i}jK}Z_j$, which imposes momentum sector selection in the DSF. Further the delta function will be replaced with a Gaussian function broadened with width $\sigma=0.03$.

In Fig.~\ref{fig:application}(b), we show the DSF at momentum $K=\pi$ for a $N=16$ site chain, with ground state depth $D=32$. Since only triplet excited states can have nonzero contributions in the DSF, the results with $n=1$ ansatz shows a good agreement with ED. In Fig.~\ref{fig:application}(c), a full energy-momentum range of the DSF with $n=1$ ansatz is shown, where the feature that DSF becomes vanishingly small with 
$K$ approaching zero
can be clearly identified. Further details for the 1D Heisenberg chain can be found in the SM.

\section{Implementation on quantum devices}
\label{sec:implementation}

In the main text and above we have shown how the basis for the tangent space excitation ansatz are constructed, for the case with and without explicit translation invariance. In computing the variational excitation spectra, the main objects we need to compute are the norm matrix and effective Hamiltonian in the variational space. Here we show how this can be achieved on quantum devices.

\begin{figure}[h]
    \centering
    \includegraphics[width=0.98\columnwidth]{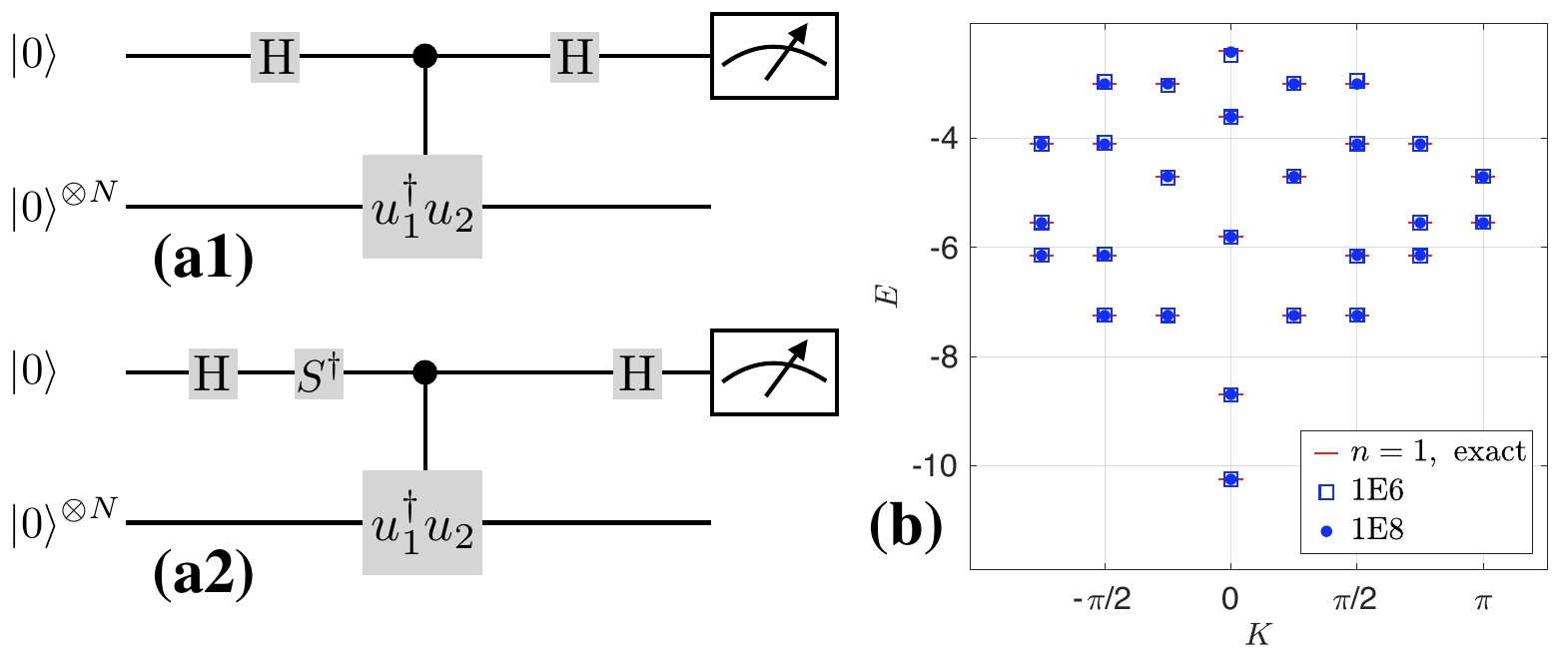}
    \caption{Hadamard test for tangent space excitation. (a1) and (a2) illustrate the circuit to measure real and imaginary part of the overlap $\langle \phi_1|\phi_2 \rangle$, respectively. Here $|\phi_j\rangle = u_j|0\rangle^{\otimes N}, j=1,2$, $\mathrm{H}$ is the Hadamard gate, and $S$ is the phase gate $S=Z^{1/2}$. Similar setup works for measuring effective Hamiltonian. In (b) we show the simulation results of Hadamard test for a $8$-site 1D TFI model at $g=1$. For simplicity, only even parity sector with $n=1$ ansatz is considered. For both $1\times 10^6$ and $1\times 10^8$ samples, one can extract the energy versus momentum with negligible error.}
    \label{fig:hadamard_test}
\end{figure}

For the case with explicit translation invariance, the basis state is given by a momentum superposition of various configurations. E.g., in 1D we have $|\phi_K(G)\rangle=\sum_{j}\mathrm{e}^{-\mathrm{i}Kj}|\phi_j(G)\rangle$, where $|\phi_j(G)\rangle$ is generated by the ground state circuit with a local perturbation gate $G$ inserted. Notice here we use $j$ to label the site index. Due to linearity, the overlap between these basis states can be classically converted to a sum of overlap for basis states without translation invariance:
\begin{equation}
    \langle \phi_K(G')|\phi_K(G)\rangle = \sum_{j,j'}\mathrm{e}^{\mathrm{i}K(j'-j)}\langle \phi_{j'}(G')|\phi_j(G)\rangle.
\label{eq:overlap_k_space}
\end{equation}
The same is true for matrix elements of effective Hamiltonian. Note that, the state $|\phi_j(G)\rangle$ is generated by a unitary acting on the initial state, and thus is easy to prepare, while $|\phi_K(G)\rangle$ takes a Bloch wave form 
and is harder to prepare.
Thus on quantum devices, we can focus on the case where the basis do not have translation invariance, and then a Fourier transform with the classical data 
generates the norm matrix and effective Hamiltonian in a given momentum sector.

The main technique to measure the overlap is Hadamard test, which is a useful tool in advanced quantum algorithms. Consider two of the basis states $|\phi_j\rangle, j=1,2$, which can be generated by a unitary circuit on an initial product state, i.e., $|\phi_j\rangle=u_j|0\rangle^{\otimes N}$. To measure $\langle \phi_1|\phi_2\rangle$, we consider the real and imaginary part separately.

For the real part, introducing an ancilla qubit as the control qubit, followed by a controlled unitary shown in Fig.~\ref{fig:hadamard_test}(a1), and finally a measurement in the $Z$ basis of the control qubit, 
the probability of the control qubit 
in the state $|0\rangle$ is given by
\begin{equation}
    P_{\mathrm{Re}}=\big(1+\mathrm{Re}(\langle 0|^{\otimes N}u_1^\dagger u_2|0\rangle^{\otimes N})\big)/2.
    \label{eq:Hadamard_test_real}
\end{equation}
To measure the imaginary part, one can further insert a phase gate in the control qubit (shown in Fig.~\ref{fig:hadamard_test}(a2)). Measuring the control qubit in the $Z$ basis, 
the probability of being in $|0\rangle$ is given by
\begin{equation}
    P_{\mathrm{Im}} = \big(1+\mathrm{Im}(\langle 0|^{\otimes N}u_1^\dagger u_2|0\rangle^{\otimes N})\big)/2.
    \label{eq:Hadamard_test_imag}
\end{equation}
Thus, combining Eqs.~\eqref{eq:Hadamard_test_real} and \eqref{eq:Hadamard_test_imag}, through measuring the control qubit, one can obtain the basis overlap $\langle \phi_1|\phi_2\rangle$. Then through a Fourier transform (Eq.~\eqref{eq:overlap_k_space}), the overlap matrix in fixed momentum sector can be obtained.

For the effective Hamiltonian, one can notice that the original Hamiltonian can be decomposed into a sum of operator strings, each term of which with basis $|\phi_j\rangle,j=1,2$ can be separately measured through the Hadamard test. And thus the effective Hamiltonian can be obtained.

Using the 1D TFI model as example, we have simulated the quantum circuits for Hadamard test classically. For simplicity, we use a $N=8$ site system at $g=1$, and only consider the $n=1$ ansatz in the even parity sector. Here the even parity sector in the excitation ansatz can be realized by keeping only the $X$ operator in the operator basis. In the main text we have shown that for this case the variational spectrum agrees well with ED. Here the goal is to see how the measurement imperfection affects the spectrum.

It turns out that, with current experimental available parameters, e.g., $1\times 10^6$ samples for each matrix element, one can reach reasonable accuracy. As shown in Fig.~\ref{fig:hadamard_test}(b), the results with samples $1\times 10^6$ and exact results show negligible difference, which can be further 
reduced by increasing the number of samples. Notice that, considering the finite precision due to measurement, the zero modes in the norm matrix become fuzzy, and one needs to introduce a threshold in the spectra of norm matrix to obtain the pseudo-inverse. The precise value of the threshold depends on the problem and also the precision. For the example shown here, it is relatively easy to identify the zero modes. Note also that, with measurement error, the excited state energy obtained is no longer variational, and can be slightly lower than exact values.

As mentioned in the main text,
the implementation on quantum devices can be simplified 
by considering the spatial translation symmetry and cancellation of parts of the unitary gates in $u_1^\dagger u_2$. A recently proposed efficient way of doing the Hadamard test has the potential to further reduce the algorithm run time~\cite{Schiffer2025}. We leave this point to future consideration.


\end{document}


\title{Supplemental Material for ``Tangent Space Excitation Ansatz for Quantum Circuits''}

\author{Ji-Yao Chen}
\email{chenjiy3@mail.sysu.edu.cn}
\affiliation{Center for Neutron Science and Technology, Guangdong Provincial Key Laboratory of Magnetoelectric Physics and Devices, School of Physics, Sun Yat-sen University, Guangzhou 510275, China}

\author{Bochen Huang}
\affiliation{Institute of Physics, Beijing National Laboratory for Condensed Matter Physics, Chinese Academy of Sciences, Beijing 100190, China}
\affiliation{School of Physical Sciences, University of Chinese Academy of Sciences, Beijing 100049, China}

\author{D. L. Zhou}
\affiliation{Institute of Physics, Beijing National Laboratory for Condensed Matter Physics, Chinese Academy of Sciences, Beijing 100190, China}
\affiliation{School of Physical Sciences, University of Chinese Academy of Sciences, Beijing 100049, China}

\author{Norbert Schuch}
\email{norbert.schuch@univie.ac.at}
\affiliation{University of Vienna, Faculty of Physics, Boltzmanngasse 5, 1090 Wien, Austria}
\affiliation{University of Vienna, Faculty of Mathematics, Oskar-Morgenstern-Platz 1, 1090 Wien, Austria}

\author{Chenfeng Cao}
\email{chenfeng.cao@connect.ust.hk}
\affiliation{HK Institute of Quantum Science $\&$ Technology, The University of Hong Kong, Hong Kong, China}
\affiliation{Dahlem Center for Complex Quantum Systems, Freie Universit\"{a}t Berlin, Berlin 14195, Germany}

\author{Muchun Yang}
\email{yang.muchun@iphy.ac.cn}
\affiliation{Institute of Physics, Beijing National Laboratory for Condensed Matter Physics, Chinese Academy of Sciences, Beijing 100190, China}
\affiliation{School of Physical Sciences, University of Chinese Academy of Sciences, Beijing 100049, China}

\date{\today}

\maketitle

To complement the main findings in the manuscript, here we provide several relevant details in this supplemental material, organized as follows. We first compare our approach with multiple existing quantum excited-state algorithms in Sec.~\ref{sec:comparison}, and then discuss fermion solution and additional data for 1D transverse field Ising (TFI) model in Sec.~\ref{sec:1D_TFI}. Hamiltonian variational ansatz (HVA) and additional tangent space excitation spectrum for 2D TFI model are presented in Sec.~\ref{sec:2D_TFI}. As further applications, HVA ansatz for spin-$1/2$ kagome Heisenberg antiferromagnet is shown in Sec.~\ref{sec:kagome}, and details of excitation spectrum and dynamical spin structure factor for spin-$1/2$ Heisenberg chain are discussed in Sec.~\ref{sec:1D_Heisenberg}. 

\section{Comparison with known quantum excited-state algorithms}
\label{sec:comparison}

In the main text we have compared the performance of our excitation ansatz with five existing quantum excited-state algorithms, finding our ansatz can capture significantly more excited states with several orders of magnitude smaller error. Here we sketch these five algorithms, discuss the related recent extensions and provide relevant implementation details.

The first algorithm we considered is the subspace expansion variational quantum eigensolver (SEVQE)~\cite{McClean2017}. Starting from the variational ground state, this algorithm constructs a subspace for the excitations by acting with local operators on the variational ground state. In the simplest case, the local operator is onsite, with which the SEVQE is equivalent to the single mode approximation and is adopted in our implementation. Further, we have used the same ground state for SEQVE as for our tangent space excitation ansatz, and found the error is smaller with both sectors starting with even parity ground state, which is the data we presented in Fig.2(f) of the main text. Note that, since the perturbation is restricted to the top layer, when implementing this algorithm in quantum devices, the matrix elements of norm matrix and the effective Hamiltonian can be obtained by measuring expectation values of local operators, and thus Hadamard test is not needed in this case.

The second algorithm is subspace search variational quantum eigensolver (SSVQE)~\cite{Nakanishi2019}. This algorithm starts with $k$ orthogonal direct product states ($|\Psi_{0,j}\rangle,j=0,1,\ldots,k-1$) as the initial states, which are evolved by the same parametrized unitary circuit $U(\bm{\theta})$ to the final states, with $\bm{\theta}$ the variational parameters. The algorithm then uses a weighted sum of expectation values of the Hamiltonian $H$ in the final states as the cost function, i.e.,
\begin{equation}
    \mathcal{L}_{w}(\bm{\theta})=\sum_{j=0}^{k-1} w_j\langle\Psi_{0,j}|U^\dagger(\bm{\theta}) H U(\bm{\theta})|\Psi_{0,j}\rangle,
\label{eq:SSVQE}
\end{equation}
where $w_j$ are the weights. With properly chosen weights, one can find both ground state and low-energy excited states simultaneously. 
This algorithm can also serve as a subroutine in hybrid quantum-classical algorithms. For instance, an application of this algorithm has been proposed in Ref.~\cite{Yalouz2021}, combining SSVQE and state averaged orbital optimization technique in quantum chemistry to render current quantum devices potentially useful for chemistry. An open source software for the combined algorithm is available in Ref.~\cite{Beseda2024}.

Since one targets the ground state and excited states at the same time, the circuit for SSVQE needs to have sufficient expressive power. In our implementation, we found that the Hamiltonian variational ansatz leads to large error in the error measure, and have used the brickwall circuit instead, where each gate is a general $\mathrm{SU}(4)$ unitary with $15$ parameters. We have further tried circuit depth $D=3N/2$ (the same as in HVA) and $D=3N$ to optimize the cost function, and found the latter provides more accurate results, which is what we have shown in the main text. In principle, this algorithm does not need Hadamard test in the implementation, since the orthogonality is preserved by the unitary circuit. In practice, we found that a diagonalization in the space formed by the final optimized states leads to smaller ground state error and preserves variational principle. This diagonalization step can be implemented with a Hadamard test or through initializing the circuit with a superposition of two different $|\Psi_{0,j}\rangle$.

The SSVQE can be integrated with other variational quantum algorithms to improve the accuracy and optimization stability. One notable example is to combine SSVQE with variational quantum imaginary time evolution~\cite{McArdle2019}, avoiding possible local minima in searching for excited states~\cite{Cianci2024}. The combined algorithm, called subspace search quantum imaginary time evolution (SSQITE) in Ref.~\cite{Cianci2024}, replaces the cost function Eq.~\eqref{eq:SSVQE} with variational quantum imaginary time evolution to optimize the variational parameters in the circuit, while the circuit architecture remains unchanged. We have implemented this algorithm for the 1D TFI model at critical point, using a brickwall circuit with depth $D=3N$, in the same manner as in SSVQE. We found that for system size $N=4$, the error measures of both SSQITE and SSVQE are in the order of $1\times 10^{-3}$, well above that of our approach.

The third algorithm is the variational quantum deflation (VQD)~\cite{Higgott2019}, which variationally computes excited states after the variational ground state $|\Psi\rangle$ was obtained. To enforce orthogonality of excited states to ground states, this algorithm has used a penalty term with sufficiently large coefficient, leading to the cost function:
\begin{equation}
    \mathcal{L}(\bm{\theta})=\langle \Phi_0|U^\dag(\bm{\theta})HU(\bm{\theta})|\Phi_0\rangle + w|\langle \Psi|U(\bm{\theta})|\Phi_0\rangle|^2,
\end{equation}
where $w$ is the weight, $|\Phi_0\rangle$ is the initial state for excitations, and $U(\bm{\theta})$ is the parametrized circuit for excitation. By adding overlaps of the next excited state with all previously converged states to the cost function, this procedure can be iterated to compute a few excited states. In our implementation, we have used the same HVA for ground state. For excited states, we found brickwall circuits have smaller error, and tested brickwall circuits with circuit depth $D=N/2$ and $D=3N/2$, the latter of which leads to smaller error and corresponds to the data shown in the main text. When implementing on quantum devices, this algorithm requires a swap test to obtain the state overlap. Similar to SSVQE, we found that in practice, to reduce error, it would be beneficial to do a diagonalization with the subspace formed by final optimized states, which requires a Hadamard test.

The fourth algorithm we have considered is the folded spectrum variational quantum eigensolver (FSVQE)~\cite{Tazi2024}, which approaches the excited states through variationally optimize the energy variance instead of energy itself. The cost function is given by:
\begin{equation}
    \mathcal{L}(\bm{\theta},\omega)=\langle\Psi_0|U^\dagger(\bm{\theta})(H-\omega)^2 U(\bm{\theta})|\Psi_0\rangle, 
\label{eq:FSVQE}
\end{equation}
where $|\Psi_0\rangle$ is an initial state, $\omega$ is a hyperparameter and $\bm{\theta}$ are the variational parameters to optimize. As the minium of Eq.~\eqref{eq:FSVQE} corresponds to the eigenstate with energy closest to $\omega$, in principle one can tune $\omega$ to reach arbitrary eigenstate. In our implementation, for excited states we have used the same circuit architecture with depth $D=3N/2$ as in VQD, while the ground state is obtained using HVA. This algorithm can be implemented in quantum devices at the cost of a squared number of terms to measure, the cost of which can be reduced using a Pauli grouping procedure~\cite{Tazi2024}. In addition, we found that a final diagonalization as in VQD can reduce error, which requires a Hadamard test.

The last algorithm we considered is the quantum Davidson algorithm (QDavidson)~\cite{Tkachenko2024}, which improves upon the quantum Lanczos algorithm~\cite{Motta2020}. Both algorithms use the quantum imaginary time evolution~\cite{Motta2020} as a subroutine, and construct a Krylov space to diagonalize the Hamiltonian. A key ingredient for this algorithm is to approximate the Trotterized imaginary time evolution with local Hamiltonian by a unitary with size of support $s$. In practice we have restricted to $s=4$ and used a second order Trotter decomposition with step $\delta\tau =0.1$ for the imaginary time evolution. The Hadamard test is needed for implementation on devices.

Comparing the performance of above methods with our approach (shown in Fig.2(f) of main text), we find that while with small system size $N=4$, VQD and QDavidson have similar accuracy with our ansatz, for larger system sizes ($N=8,12,16$), our approach significantly improves the accuracy over the five methods, reducing error by several orders of magnitude. Thus, we believe our approach represents a state-of-the-art advancement
for quantum excited-state algorithms.


\section{Details of excitations of 1D TFI model}
\label{sec:1D_TFI}

In this section, we discuss more details of the tangent space excitation of 1D transverse field Ising model, and compare the variational spectrum to its fermion solution. This model is a textbook example of exactly solvable models in quantum many-body physics, and the fermion solution has been known for several decades. The solution we present below largely follows Ref.~\cite{Mbeng2024}, which we will go through briefly and only discuss the relevant results.

\subsection{Exact fermion solution}

In the main text, the Hamiltonian reads $H = -\sum_{i}Z_iZ_{i+1} + g X_i$, with periodic boundary condition (PBC) imposed. Here and below, we only consider the system size $N$ even case. To solve this model using Jordan-Wigner transformation, we first carry out an onsite unitary transformation to exchange $Z$ and $X$, resulting in the following Hamiltonian
\begin{equation}
    H = -\sum_{i} X_iX_{i+1} + g Z_i,
\label{eq:1D_TFI}
\end{equation}
which shares the same spectrum as the original one. This is the starting point of the following analysis.

Under the Jordan-Wigner transformation, one can map the Hamiltonian Eq.~\eqref{eq:1D_TFI} to a spinless free fermion model. Namely, introducing one spinless fermion mode with creation (annihilation) operator $f^\dagger$ ($f$) for each site, we can represent the spin operators using fermion operators:
\begin{equation}
    Z_i = 2f_i^\dagger f_i -1,\ X_i = (f_i^\dagger + f_i )\mathrm{exp}(i\pi \sum_{j<i} n_j),
\end{equation}
where a string operator is needed to ensure the correct commutation relations in fermion operators. The fermion model after the mapping is quadratic, taking a form:
\begin{equation}
\begin{split}
    H =& -\sum_{i=1}^{N-1}(f_i^\dagger - f_i)(f_{i+1}^\dagger + f_{i+1}) - g \sum_{i=1}^N (2f_i^\dagger f_i - 1) \\
    & + Q(f_N^\dagger - f_N)(f_1^\dagger + f_1),
\end{split}
\label{eq:1D_fermion}
\end{equation}
where $Q=\mathrm{exp}(i\pi\sum_{j=1}^N n_j)$ is the fermion parity operator. Note that, the last term in Eq.~\eqref{eq:1D_fermion} appears due to the PBC in the spin Hamiltonian. This Hamiltonian does not conserve fermion number, and involves fermion pairing instead.

As one can see from Eq.~\eqref{eq:1D_TFI}, the $\mathbb{Z}_2$ symmetry is generated by a product of Pauli $Z$, which after the mapping becomes the fermion parity operator. Specifically, the $\mathbb{Z}_2$ even sector corresponds to the fermion parity even sector, and the $\mathbb{Z}_2$ odd sector is mapped to the fermion parity odd sector. Crucially, the fermion model Eq.~\eqref{eq:1D_fermion} depends on the fermion parity explicitly. For even fermion parity, we have $Q=1$ and thus Eq.~\eqref{eq:1D_fermion} takes a anti-periodic boundary condition (ABC). For odd fermion parity, we have $Q=-1$ and Eq.~\eqref{eq:1D_fermion} takes the periodic boundary condition instead.

With different boundary conditions, the fermion modes in momentum space will have different allowed momenta. When the fermion model takes ABC, the allowed momentum quantum number is given by $k=2\pi/N\times (m+1/2)$, $m=0,1,2,...,N-1$. When the fermion model takes PBC, the allowed momenta is $k=2\pi/N \times m$, $m=0,1,2,...,N-1$.

Diagonalizing the quadratic fermion model Eq.~\eqref{eq:1D_fermion} in momentum space, one can find that, in the even parity sector, the ground state is given by a Bardeen-Cooper-Schrieffer (BCS) fermion pairing form, with the energy of a Bogoliubov mode with momentum $k$ (or $-k$) given by 
\begin{equation}
\epsilon_k = 2\sqrt{(g-\mathrm{cos}(k))^2+\mathrm{sin}^2(k)}.
\label{eq:Bogoliubov_spec}
\end{equation}
The excited states in this sector also have to satisfy the even parity constraint, which are generated by applying even number of creation operators of Bogoliubov modes to the ground state (i.e., the vacuum of Bogoliubov modes). Denote the total occupation number in the Bogoliubov modes as $n_B$. The ground state in even sector then has $n_B=0$, while excited states will have $n_B=2,4,6,...$. The latter correspond to the ``2 fermion'' states, ``4 fermion'' states, etc, mentioned in Fig.2 of the main text.

In the odd parity sector, the fermion modes at momentum $k=0$ and $k=\pi$ are not paired up, and in fact decouple from other $k$ modes. To satisfy the fermion parity odd constraint, the ground state should have one extra fermion occupied in the $k=0$ or $k=\pi$ mode in addition to the BCS pairing state for other $k$ modes. With our choice of parameters, in the ground state, the $k=0$ fermion mode is occupied, while $k=\pi$ mode is empty. The energy of a Bogoliubov mode with momentum $k$ (or $-k$) is still given by Eq.~\eqref{eq:Bogoliubov_spec}, while the energy of fermion mode $k=0$, $k=\pi$ is given by $\epsilon_0=-2+2g$, $\epsilon_{\pi}=2+2g$, respectively.

Due to the occupied $k=0$ fermion mode in the ground state, the excited states in the odd sector have a rich structure. Denote the occupation number in $k=0$ and $k=\pi$ fermion mode as $n_0$ and $n_{\pi}$, respectively. The many-body state in the odd sector can be labeled by the tuple $(n_B, n_0, n_{\pi})$. As the discussed above, the ground state in odd sector has $(n_B,n_0,n_{\pi})=(0,1,0)$. On top of that, the low energy excitation would be two fermion excited states, which can be labeled as $(n_B,n_0,n_{\pi})=(1,0,0), (0,0,1)$. This is the ``1 fermion'' state mentioned in the main text. Here we term them ``1 fermion'' state due to $n_B+n_0+n_{\pi}=1$. Similarly, we can also have $(n_B,n_0,n_{\pi})=(2,1,0),(1,1,1),(2,0,1),(3,0,0)$ for higher energy excited states, which we term as ``3 fermion'' state in the main text. Note that, numerically we found that at the critical point ($g=1$), the first two cases in the ``3 fermion'' state can be captured by the tangent space excitation ansatz with $n=1$, while the latter two are completely missed. With this distinction, we have termed the former as type-I 3 fermion state (in short, ``3 fermion, I''), and the second case as type-II (``3 fermion, II''). Similar analysis also applies to larger number of occupied modes.

The momentum quantum number of the many-body state can be obtained in a similar way. Notice that, for both even and odd sectors, the vacuum of the Bogoliubov mode is translation invariant with momentum $K=0$. (We use $k$ for momentum of fermion mode or Bogoliubov mode, and $K$ for momentum of many-body state.) Therefore, the ground state of both sectors have total momentum $K=0$. For excited states, one can simply add the momentum of each occupied Bogoliubov mode (and fermion mode in the odd sector), to obtain the total momentum.

\begin{figure}[h]
\centering
    \includegraphics[width=0.95\columnwidth]{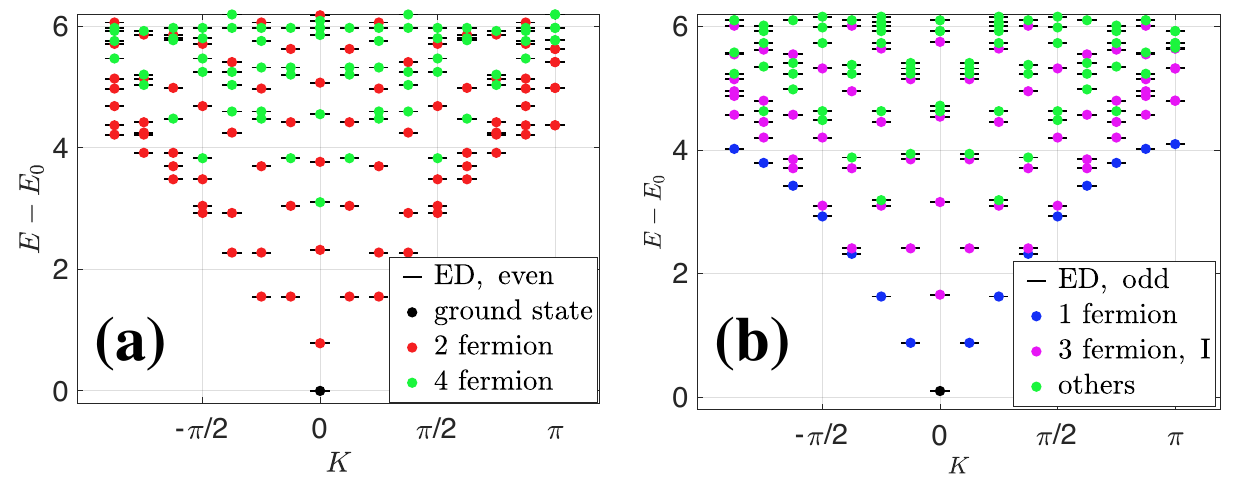}
\caption{Comparison between ED and fermion solution of the 1D TFI model. The system size is $N=16$, and Hamiltonian parameter $g=1$. In both (a) and (b), $E_0$ is the ground state energy in the even sector.}
\label{fig:1D_TFI_fermion}
\end{figure}

To make sure that the fermion solution discussed above is correct, we have carried out a self-consistence check with exact diagonalization of the spin model Eq.~\eqref{eq:1D_TFI}. The results are shown in Fig.~\ref{fig:1D_TFI_fermion}, finding perfect agreement between the two methods.

\subsection{Additional data for tangent space excitation}

\begin{figure*}[htb]
\centering
\includegraphics[width=1.95\columnwidth]{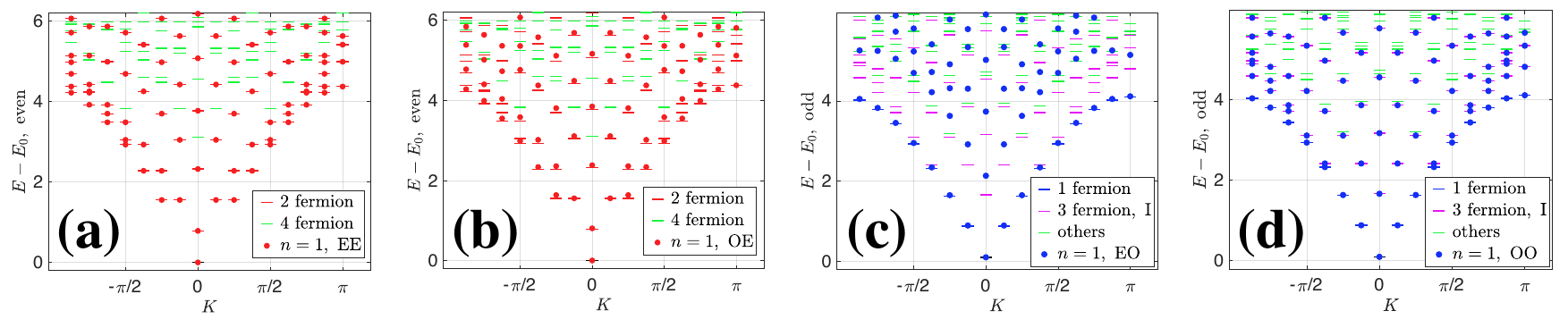}
\caption{Comparison of $n=1$ excitation spectra with two different initial states at critical point $g=1$ and system size $N=16$. For (a),(c) the initial state is $|\Psi_0\rangle=|+\rangle^{\otimes N}$, while for (b),(d) the initial state is $|\Psi'_0\rangle=\frac{1}{\sqrt{2}}\big(|0\rangle^{\otimes N} - |1\rangle^{\otimes N} \big)$. The former initial state is parity even, and through optimizing the circuit parameters we are targeting the true ground state. The latter is parity odd, and the variational parameters in the circuit are separately optimized to target the first excited state in the many-body spectrum. In both cases, the obtained excitation spectra can be further labeled by parity of each excited states. The spectra in parity even sector is shown in (a),(b), while that with parity odd sector is shown in (c),(d). In all plots, $E_0$ refers to true ground state energy.}
\label{fig:1D_TFI_n_1}
\end{figure*}

\begin{figure*}[htb]
\centering
\includegraphics[width=1.95\columnwidth]{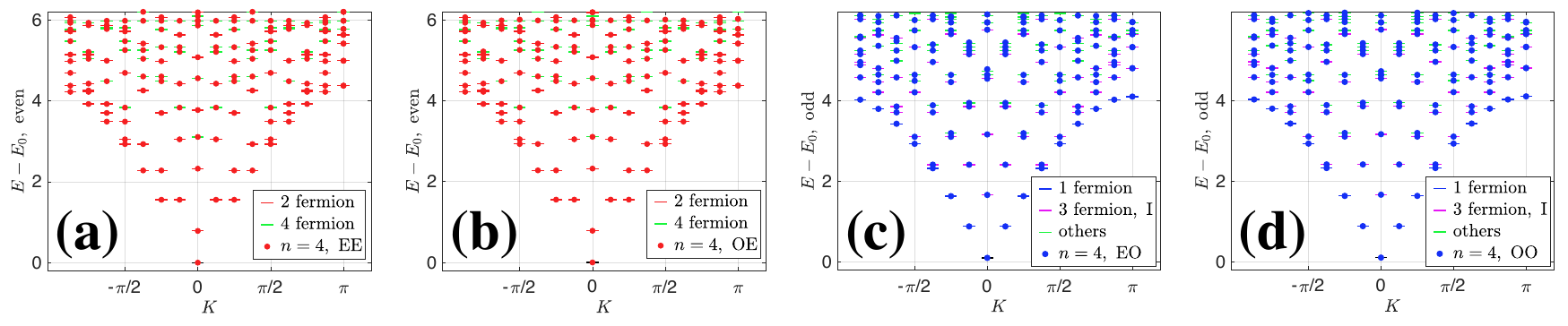}
\caption{Same setup as Fig.~\ref{fig:1D_TFI_n_1}, with the $n=4$ ansatz. Not only more states can be captured, but also the difference between spectra with different initial states becomes smaller.}
\label{fig:1D_TFI_n_4}
\end{figure*}

In this subsection, we first compare the tangent space excitation of $n=1$ ansatz with different initial states in both sectors. The resulting variational spectra will be labeled as ``EE'', ``EO'', ``OE'' and ``OO'', where ``EE'' represents variational spectrum in even parity sector starting with even parity initial state (i.e., $|\Psi_0\rangle=|+\rangle^{\otimes N}$), ``EO'' represents variational spectrum in odd parity sector using the same initial state, ``OE'' stands for variational spectrum in even parity sector starting with odd parity GHZ type initial state $|\Psi'_0\rangle=\frac{1}{\sqrt{2}}\big(|0\rangle^{\otimes N} - |1\rangle^{\otimes N} \big)$, and ``OO'' means variational spectrum in odd parity sector using odd parity initial state $|\Psi_0'\rangle$. Note that, ``EE'' and ``EO'' share the same optimized variational parameters, and similarly for ``OE'' and ``OO''. In the following, we will focus on Hamiltonian parameter $g=1$, the result of which is representative for all $g$ values.

For Hamiltonian parameter $g=1$ with system size $N=16$, using a circuit depth $D=3N/2$, we first obtain the variational ground state and then compute the excitation spectra. The results are shown in Fig.~\ref{fig:1D_TFI_n_1}. Comparing Fig.~\ref{fig:1D_TFI_n_1}(a) and (b), one can see clearly that for the even parity sector, the results with initial state parity even has better accuracy than that with odd parity initial state. Similarly, for the odd parity sector, the results with odd parity initial state has a better accuracy than that with even parity initial state, as shown in Fig.~\ref{fig:1D_TFI_n_1}(c) and (d). Nevertheless, this difference between initial states becomes small as we increase $n$. For instance, see Fig.~\ref{fig:1D_TFI_n_4} with $n=4$ where all other parameters are kept the same as $n=1$.


As mentioned in the main text, another prominent feature of increasing $n$ is that more states can be captured. {One can see in Fig.~\ref{fig:1D_TFI_n_1}(a), certain low-energy levels are not captured by the $n=1$ excitation ansatz, while levels with higher energy can be approximated to a high accuracy. To figure out the precise nature of these energy levels, we have compared the results with the exact fermion solution presented above. We find that the 2 fermion excitation can be captured by the $n=1$ ansatz, while 4, 6 fermion excitations are completely missed. Similarly, for the odd parity sector (see Fig.~\ref{fig:1D_TFI_n_1}(d)), only the states labeled as “1 fermion” states and “3 fermion, I” states can be captured. This behavior can be understood from the fact that the whole circuit maps to free fermions and the $n=1$ ansatz contains at most two fermionic operators. With increasing $n$, the missing levels can be captured well.}
See comparison between Fig.~\ref{fig:1D_TFI_n_1} and Fig.~\ref{fig:1D_TFI_n_4}, where ``4 fermion'' states in even parity sector and ``3 fermion, II'' states in odd sector can be captured with $n=4$. 
{However, as the tangent space dimension scales exponentially with the order $n$, $n$ has to be limited to a small value.}
It would be interesting to find a more efficient way of representing multi-particle excitations in quantum circuits.

\begin{figure}[h]
\centering
\includegraphics[width=0.95\columnwidth]{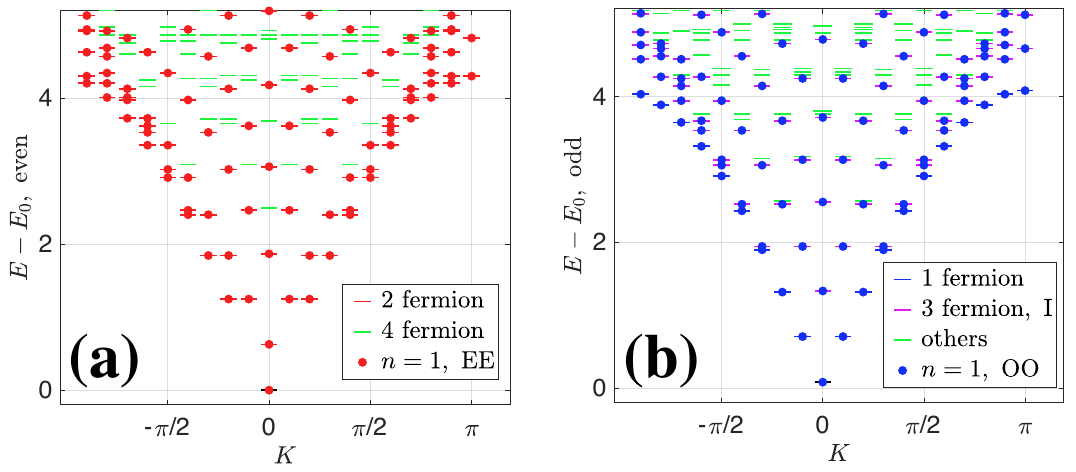}
\caption{Tangent space excitation spectrum for 1D TFI model with system size $N=20$ at $g=1$. Here we use the $n=1$ ansatz with ground state circuit depth $D=3N/2$. (a) shows the spectrum in even parity sector with initial state $|\Psi_0\rangle=|+\rangle^{\otimes N}$, and (b) is for the spectrum in odd parity sector with initial state $|\Psi'_0\rangle=\frac{1}{\sqrt{2}}\big(|0\rangle^{\otimes N} - |1\rangle^{\otimes N} \big)$. In both (a) and (b), $E_0$ refers to true ground state energy.}
\label{fig:1D_TFI_N_20}
\end{figure}

As we have shown in Fig.2(e) of main text, the error of single particle excitation (in spin language, comparing to ED results) is on the order of $1\times 10^{-11}$, and this precision is in fact limited by the optimized ground state. This small error suggests that our approach is scalable, at least for the single particle excitation. Here we provide more numerical evidence for the scalability. In Fig.~\ref{fig:1D_TFI_N_20}, we show the $n=1$ excitation spectrum for the $N=20$ case in both even and odd parity sector, starting from initial state with even and odd parity, respectively. Similar to the $N=16$ case, for the even parity sector, all the ``2 fermion'' states can be captured to a high precision, while for the odd sector, the ``1 fermion'' and ``3 fermion, I'' states can be captured.

\begin{figure}[h]
    \centering
    \includegraphics[width=0.95\columnwidth]{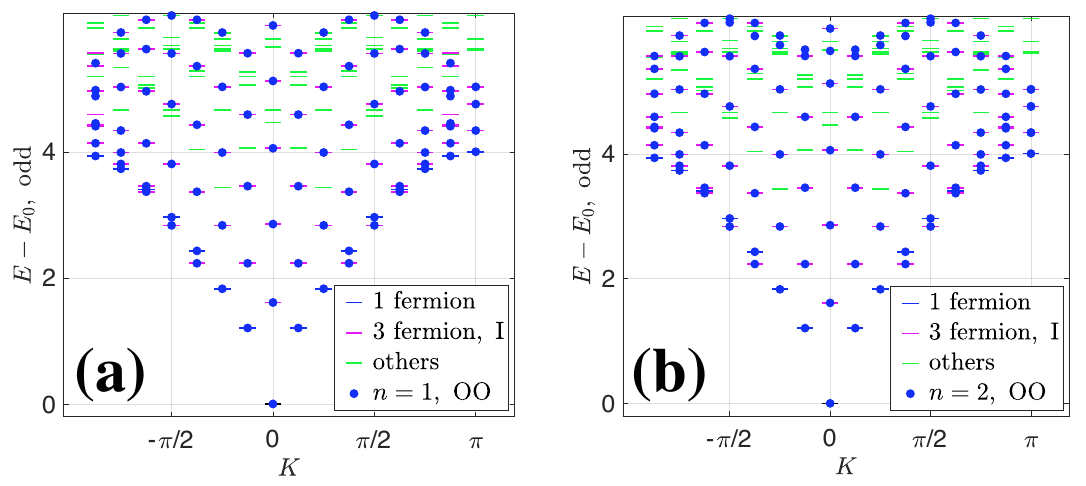}
    \caption{Comparison of $n=1$ and $n=2$ ansatz in the odd parity sector for $g=0.8$. Small deviations exist in (a) for high energy levels at momentum $K=\pm 7\pi/8$ for ``3 fermion, I'' state, which can be eliminated with $n=2$ shown in (b). Here $E_0$ is the true ground state energy.}
    \label{fig:1D_TFI_g_0.8}
\end{figure}

Before closing this section, we would like to mention that the expressive power of $n=1$ ansatz with even parity and odd parity initial state may not be the same. Indeed, we have found numerically that deep in the ferromagnetic phase, while ``2 fermion'' state can always be captured by $n=1$ even ansatz, the high energy part of the ``1 fermion'' and ``3 fermion, I'' state can have small error with $n=1$ odd ansatz. See Fig.~\ref{fig:1D_TFI_g_0.8} (a) for an example at $g=0.8$. An intuitive explanation is that away from the critical point, the ground state in odd sector is close to the initial state $|\Psi_0\rangle=\frac{1}{\sqrt{2}}\big(|0\rangle^{\otimes N} - |1\rangle^{\otimes N} \big)$. Thus the unitary circuit would be less effective in generating entanglement (or being close to identity). Therefore adding perturbation layer with $n=1$ would have limited power in constructing excited states. Nevertheless, one can simply increase from $n=1$ to $n=2$ to capture the ``1 fermion'' and ``3 fermion, I'' state in odd parity spectrum, as shown in Fig.~\ref{fig:1D_TFI_g_0.8}(b). This picture in fact is also in line with the physical heuristics based on adiabatic evolution mentioned in the main text.


\section{More details for 2D TFI model}
\label{sec:2D_TFI}

\begin{figure}[h]
    \centering
    \includegraphics[width=0.95\columnwidth]{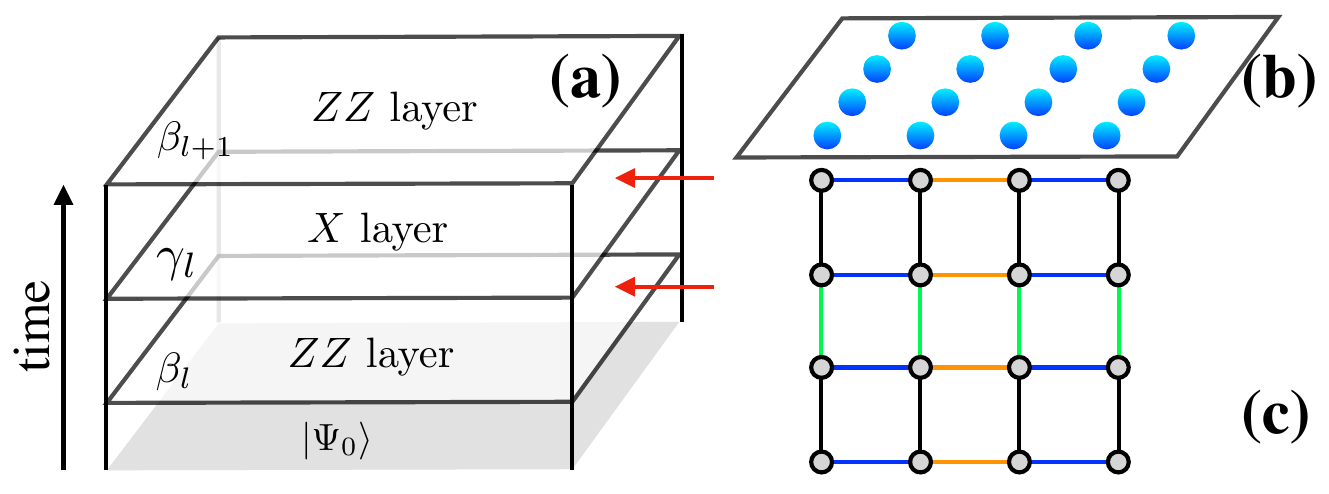}
    \caption{HVA for 2D TFI model. In (a), the circuit for variational ground state is composed by layers of $ZZ$ gate and $X$ gate, where each $ZZ$ layer shares the same variational parameter $\beta_l$ and each $X$ layer has variational parameter $\gamma_l$, with $l$ the layer index. The $X$ layer is given by a product of one qubit gate, taking a form $U_X=\mathrm{exp}(-\mathrm{i}\frac{\gamma}{2}X)$, shown in (b). The $ZZ$ layer can be further decomposed into four layers where each layer is given by a product of $ZZ$ gate acting on non-overlapping nearest-neighbor qubits, shown as colored links in (c). Here each $ZZ$ gate takes the form $U_{ZZ}=\mathrm{exp}(-\mathrm{i}\frac{\beta}{2}Z_iZ_j)$ for qubits $i,j$. The red arrow in (a) indicates the time location where we could insert an extra layer to construct the excitation basis, for each of which, only one layer is inserted.}
    \label{fig:2D_Ising_circuit}
\end{figure}

Here we show more details of simulations for the 2D transverse field Ising model. 
A schematic figure of the HVA for the 2D TFI model is shown in Fig.~\ref{fig:2D_Ising_circuit}. For the system size $N=4\times 4$ considered in the main text, there are eight $ZZ$ layers and eight $X$ layers in the HVA circuit, where each of them has an independent parameter. Thus the circuit depth is $D=5\times 8=40$, and the number of variational parameters is $8+8=16$.

\begin{figure}[h]
\centering
\includegraphics[width=0.95\columnwidth]{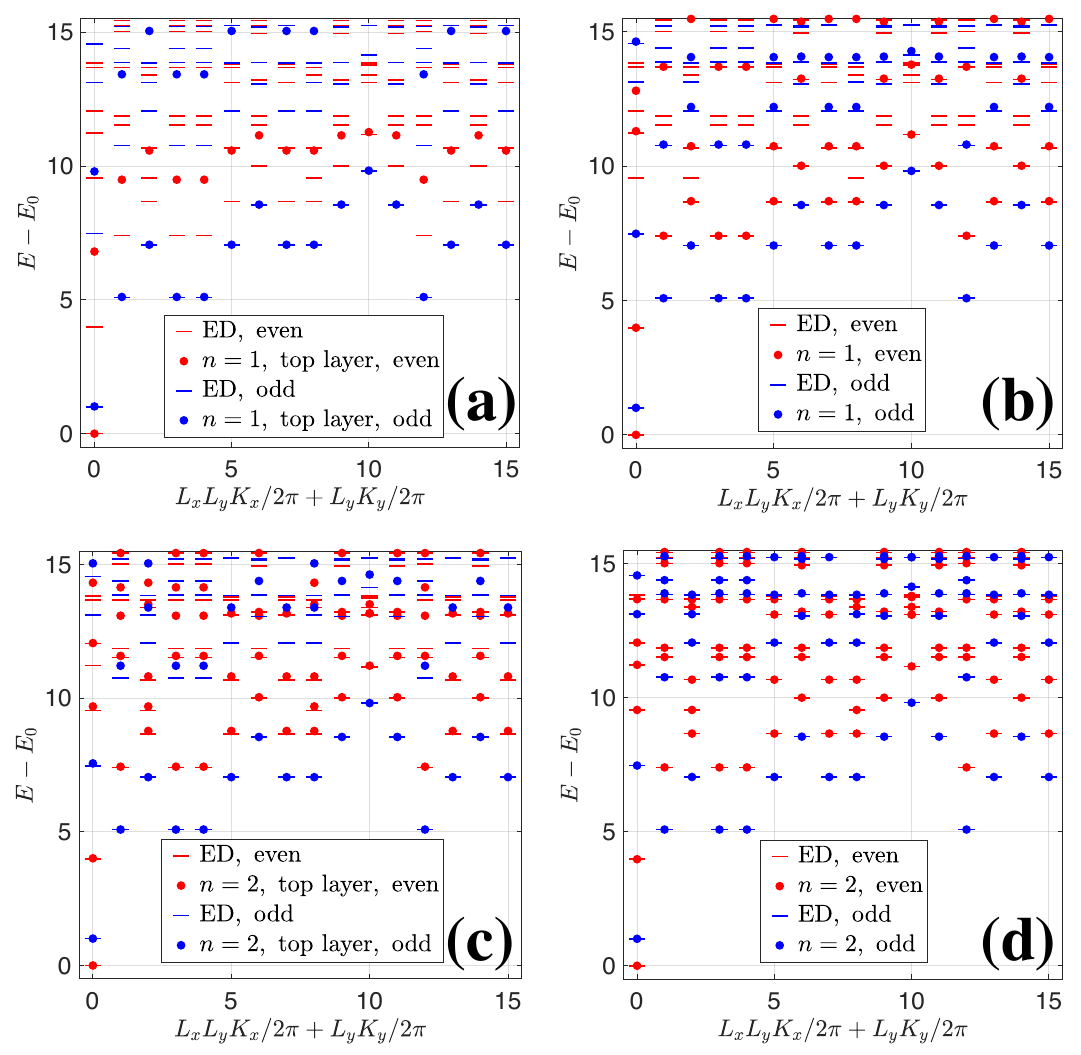}
\caption{Comparison of variational spectra with ED at $g=3$ for 2D TFI model. In (a) and (c), the perturbation layer is only on the top, and in (b) and (d) we have included all basis. (a) is equivalent to the SMA approach. In all figures, $E_0$ is the true ground state energy.}
\label{fig:2D_Ising_comp}
\end{figure}

As mentioned in the main text, for the 2D TFI model, the $n=2$ ansatz is slightly better than the $n=1$ ansatz. In Fig.~\ref{fig:2D_Ising_comp} we compare the single mode approximation, $n=1$, and $n=2$ ansatz, finding that while $n=1$ and $n=2$ ansatz have similar performance for the lowest excitation in each momentum sector, the $n=2$ ansatz has better accuracy at higher excitations.


\section{Simulation details of spin-$1/2$ kagome Heisenberg antiferromagnet}
\label{sec:kagome}

In this section, we present the details of the ground state optimization for kagome Heisenberg model, and the basis construction for tangent space excitation.

\begin{figure}[h]
\centering
\includegraphics[width=0.95\columnwidth]{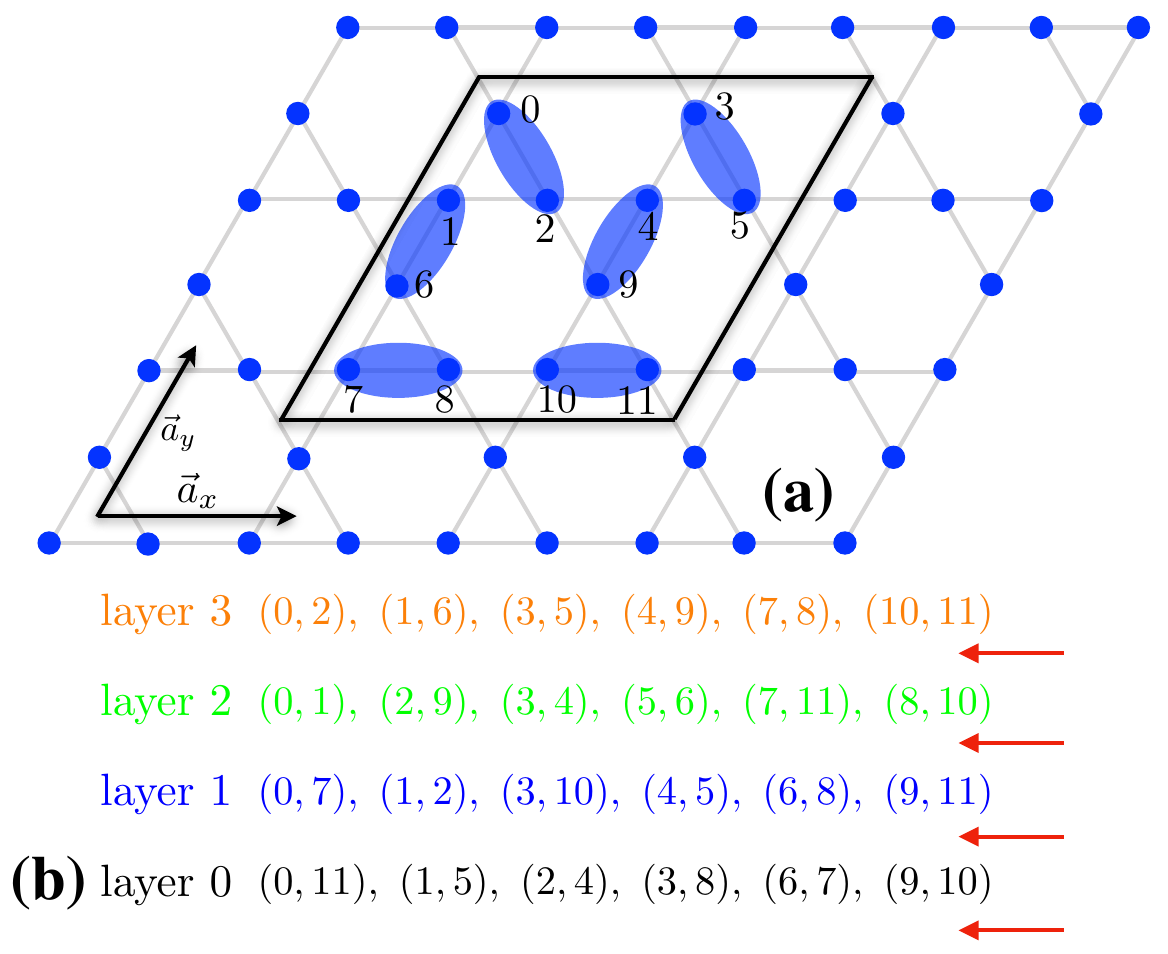}
\caption{HVA for kagome Heisenberg model. (a) The kagome lattice is generated by translation along $\vec{a}_x=(1,0)^{\mathrm{T}}$, $\vec{a}_y=(1/2,\sqrt{3}/2)^{\mathrm{T}}$ directions with a three-site unit cell. The $12$-site cluster studied in this work is indicated by the parallelogram, where PBC along $\vec{a}_x$ and $\vec{a}_y$ directions are imposed. The initial state for the HVA ansatz is indicated by the blue ovals, each of which represents a spin singlet. (b) The ground state circuit is given by repeated actions of layer $0,1,2,3$ to the initial state, where the support of gates in each layer is indicated by the site label. 
The red arrow indicates the time location we insert perturbation layer to construct excitation basis.}
\label{fig:kagome_schematic}
\end{figure}

Following Ref.~\cite{Kattemolle2022}, we use a HVA ansatz native to the kagome lattice for the spin-$1/2$ kagome Heisenberg model, shown in Fig.~\ref{fig:kagome_schematic}. The $12$-site cluster we studied is indicated by the parallelogram. Since the ground state is a spin singlet, we use a valence bond covering (shown as blue ovals) as the initial state, where each valence bond represents a spin singlet. The variational ground state is then obtained by repeatedly applying layer $0,1,2,3$ to the initial state, with a circuit depth $D=4\times 16=64$. All gates in each layer are generated by time evolution with a local Heisenberg Hamiltonian, taking a form $U=\mathrm{exp}(-i\beta/2(X_iX_j+Y_iY_j+Z_iZ_j))$ acting on qubits $i,j$, which is parametrized by the angle $\beta$. Since the initial state and the circuit structure breaks the translation symmetry explicitly, all parameters $\beta$'s are taken to be independent.

To construct the basis for tangent space excitation, we insert a new layer with the time location indicated by red arrows in Fig.~\ref{fig:kagome_schematic}(b). The locality of the inserted gate follows the kagome lattice geometry. Since there is no explicit translation symmetry in the ground state circuit, we do not impose translation symmetry in the excited state basis either. Thus each insertion of a perturbation gate generates a new basis, the collection of which gives the space for the excitations. Here we would like to mention that, although translation symmetry is not explicit, similar to the spin-$1/2$ Heisenberg chain in section Sec.~\ref{sec:1D_Heisenberg}, we have checked that the variational ground state and excited states are translationally invariant. And thus momentum quantum number can be meaningfully obtained. The results have been shown and discussed in the main text.


\section{Details of excitations for 1D Heisenberg model}
\label{sec:1D_Heisenberg}

To show the ``usefulness'' of the tangent space excitation ansatz, we have studied the dynamical spin structure factor (DSF) of the spin-$1/2$ Heisenberg chain, which plays an important role in quantum magnetism and many-body physics. In the End Matter we have shown the DSF results with the excitation ansatz, here we discuss the details of the variational ground state circuit and the excitation ansatz.

The Hamiltonian of this model is defined as:
\begin{equation}
    H = \sum_{i=0}^{L-1} X_iX_{i+1} + Y_iY_{i+1} + Z_i Z_{i+1},
\label{eq:1D_Heisenberg}
\end{equation}
where 
PBC is imposed with $0\equiv L$. The model has a global $\mathrm{SU}(2)$ spin rotation symmetry, and the system size we considered is $N=16$, for which it is known that the ground state is a total spin singlet. Under PBC, the eigenstates can be labeled by momentum and total spin quantum number.

\begin{figure}[h]
    \centering
    \includegraphics[width=0.95\columnwidth]{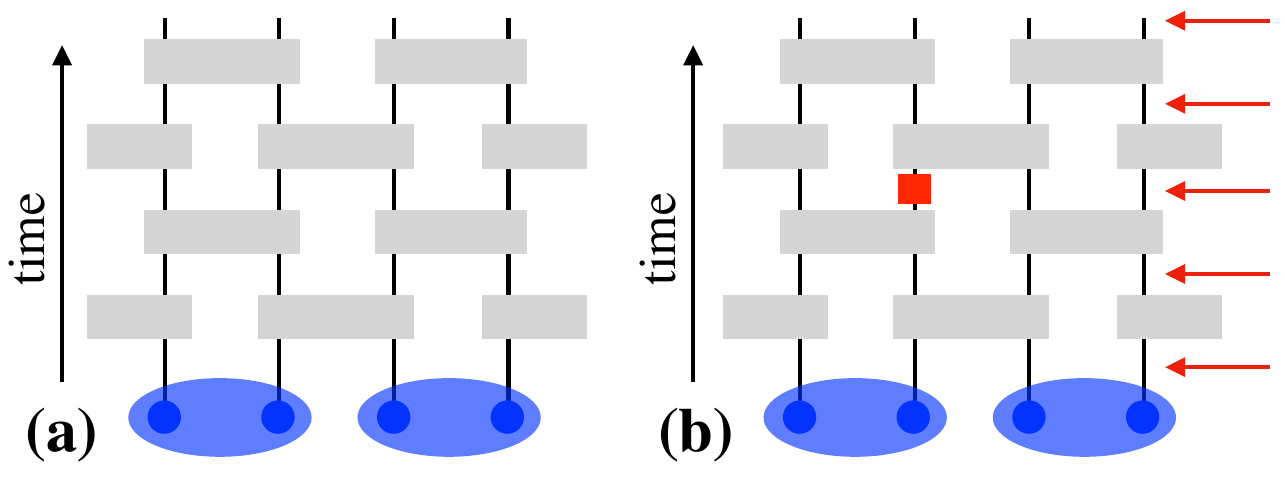}
    \caption{Tangent space excitation of spin-$1/2$ Heisenberg chain. (a) shows the HVA for ground state, where the parameters in each gate vary independently. The blue oval represents the singlet formed by two neighboring spin-$1/2$s. In (b) we show the basis states for the excitation ansatz, with the time location of the added gate indicated by the red arrow. Note that, only a single gate is added in one basis (shown as a red square for a $n=1$ ansatz).}
    \label{fig:1D_Heisenberg_schematic}
\end{figure}

To find the variational ground state, we use a HVA shown in Fig.~\ref{fig:1D_Heisenberg_schematic}(a). To ensure that the variational ground state is a spin singlet, we use a nearest-neighbor dimer covering of the 1D chain as the initial state, where each dimer represents the two spins being in a singlet, i.e., $|\Psi_0\rangle =\otimes_{i=0}^{N/2-1}\frac{1}{\sqrt{2}}\big(|01\rangle-|10\rangle\big)_{2i,2i+1}$. The unitary gate in the circuit is given by the time evolution operator of local Hamiltonian, i.e.,
$
U(\beta) = \mathrm{exp}\big\{ -i\beta(X_{i}X_{i+1}+Y_iY_{i+1}+Z_iZ_{i+1})/2 \big\},
$
where $\beta$ is the variational parameter to optimize.
Different from the HVA used for the TFI models, here we treat all gate parameters independently, and thus the variational ground state does not have explicit translation invariance. This choice is motivated by the fact that the initial state breaks the one-site translation invariance explicitly. Nevertheless, one can expect that, with sufficient accuracy of the variational energy, the variational ground state is still translationally invariant to a high precision. This is indeed what we find numerically.

For the $N=16$ site chain, we find that using a circuit depth $D=2N$, the variational energy is $E_0=-28.5689$, with an error $2.6\times 10^{-4}$ comparing to the exact value. And the expectation value of one site translation operator is $1-1.7\times 10^{-5}$, ensuring that translation symmetry is restored in the final state.

With the optimized circuit available, we can now add a perturbation layer to compute the tangent space excitations. Since the unitary in each layer does not commute with translation operator explicitly, it is more reasonable to keep all the basis before momentum superposition in one space, without enforcing the translation invariance. A typical basis state is shown in Fig.~\ref{fig:1D_Heisenberg_schematic}(b). Then we diagonalize the Hamiltonian in this space to obtain the excitation spectra.

As mentioned in the main text, the $n=1$ ansatz only captures the triplet excitation, while the $n=2$ ansatz can reveal both singlet and triplet excitations. A crucial observable for quantum magnetism is the triplet gap and the singlet gap, which is defined by the energy of the first triplet (singlet) excitation relative to the ground state energy, respectively. Numerically, using the ground state energy from ED as the reference,  we find that, the triplet gap obtained from $n=1$ ($n=2$) ansatz is 1.0815 (1.0810), while the singlet gap obtained from $n=2$ ansatz is 1.7832. As a comparison, the triplet (singlet) gap from ED is 1.0808 (1.7830), showing that our ansatz indeed has a high accuracy.

\begin{figure}[h]
    \centering
    \includegraphics[width=0.95\columnwidth]{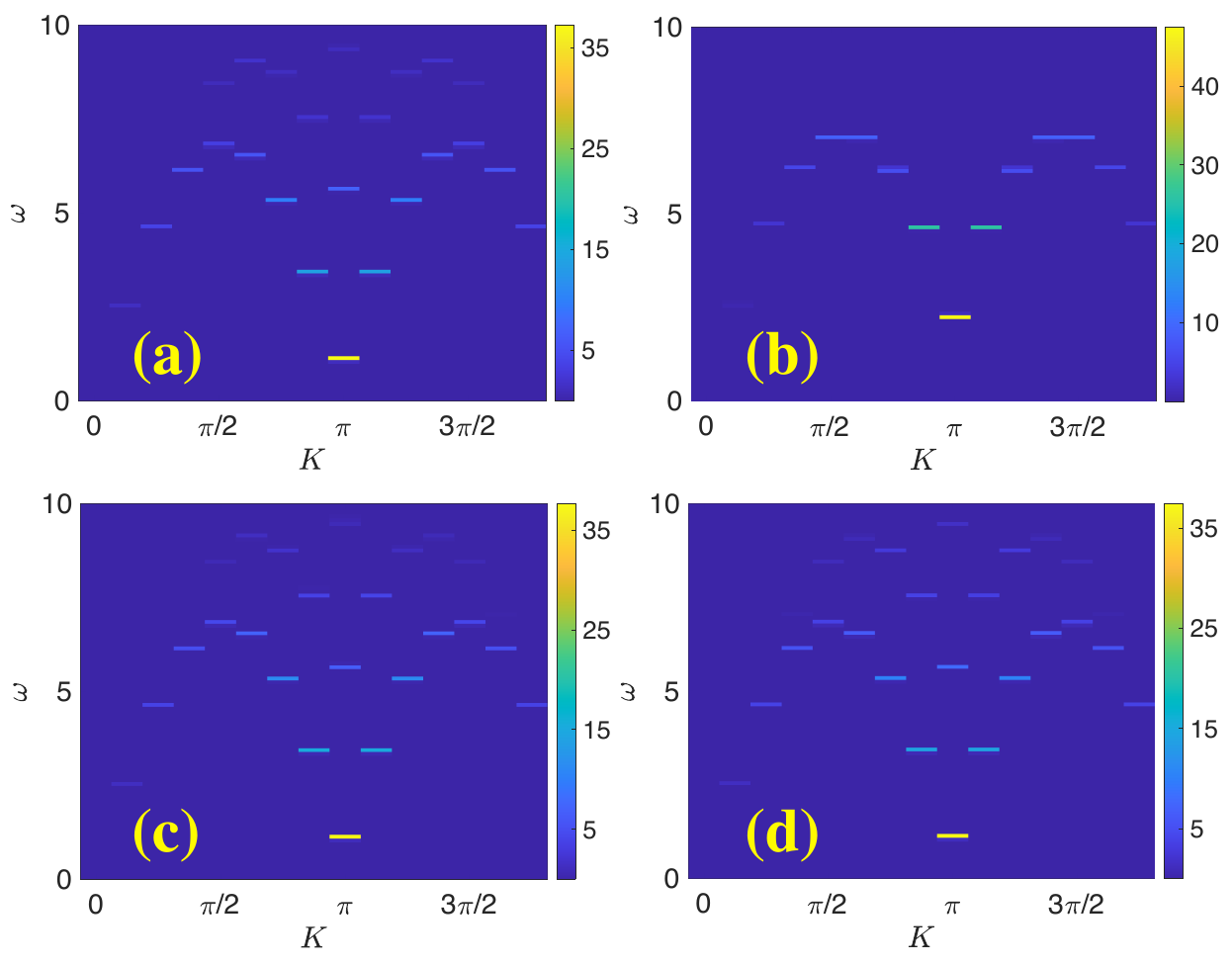}
    \caption{DSF of a $16$-site spin-$1/2$ Heisenberg chain. (a) DSF computed with ED.
    (b) DSF computed with SMA. (c) and (d) show DSF from $n=1$ and $n=2$ excitation ansatz, respectively.}
    \label{fig:1D_Heisenberg_DSF}
\end{figure}

In Fig.~\ref{fig:1D_Heisenberg_DSF}, we further compare the DSF computed with SMA, $n=1$ and $n=2$ ansatz with that from ED, showing that our ansatz significantly improves the SMA, and finding good agreement with ED. Moreover, comparing the $n=1$ and $n=2$ ansatz with that from ED, one can find that the improvement of larger $n$ is actually insignificant for DSF. Given the much smaller space dimension with $n=1$, this suggests a form of quantum advantage may be achieved with a simple $n=1$ ansatz.

\bibliography{draft}